\documentclass[draftclsnofoot, onecolumn, doublespace, 12pt]{IEEEtran}
\usepackage{epsfig,amsmath,graphics,amssymb,fancyhdr,cite}

\begin{document}
\pagestyle{fancy}
\renewcommand{\headrulewidth}{0pt} 
\lhead{\scriptsize SUBMITTED TO IEEE JOURNAL ON SELECTED AREAS IN 
  SIGNAL PROCESSING.}
\rhead{\scriptsize \thepage}
\cfoot{}

\title{Scaling Laws of Cognitive Networks}

\author{Mai Vu,$^{1}$
Natasha Devroye,$^{1}$, Masoud Sharif,$^{2}$ and Vahid Tarokh$^{1}$ \\
$^{1}$\textit{Harvard University, e-mail: maivu, ndevroye, vahid@seas.harvard.edu}\\
 $^{2}$\textit{Boston University,  e-mail: sharif@bu.edu}}
\newtheorem{theorem}{Theorem}
\newcommand{\bh}{{\bf h}}
\newcommand{\bg}{{\bf g}}
\newcommand{\bx}{{\bf x}}
\newcommand{\bs}{{\bf s}}
\newcommand{\by}{{\bf y}}
\newcommand{\bz}{{\bf z}}
\newcommand{\bn}{{\bf n}}
\newcommand{\bm}{{\bf m}}
\newcommand{\bv}{{\bf v}}
\newcommand{\bu}{{\bf u}}
\newcommand{\bw}{{\bf w}}
\newcommand{\bc}{{\bf c}}
\newcommand{\bd}{{\bf d}}
\newcommand{\bq}{{\bf q}}
\newcommand{\bH}{{\bf H}}
\newcommand{\bR}{{\bf R}}
\newcommand{\bX}{{\bf X}}
\newcommand{\bC}{{\bf C}}
\newcommand{\bF}{{\bf F}}
\newcommand{\bA}{{\bf A}}
\newcommand{\bW}{{\bf W}}
\newcommand{\bU}{{\bf U}}
\newcommand{\bV}{{\bf V}}
\newcommand{\bD}{{\bf D}}
\newcommand{\bI}{{\bf I}}
\newcommand{\bQ}{{\bf Q}}
\newcommand{\bB}{{\bf B}}
\newcommand{\bG}{{\bf G}}
\newcommand{\bE}{{\bf E}}
\newcommand{\bY}{{\bf Y}}
\newcommand{\bN}{{\bf N}}
\newcommand{\bS}{{\bf S}}
\newcommand{\bZ}{{\bf Z}}
\newcommand{\bPhi}{{\bf \Phi}}
\newcommand{\bLambda}{{\bf \Lambda}}
\newcommand{\blambda}{{\bf \lambda}}
\newcommand{\bGamma}{{\bf \Gamma}}
\newcommand{\bSigma}{{\bf \Sigma}}
\newcommand{\bTheta}{{\bf \Theta}}
\newcommand{\bPsi}{{\bf \Psi}}
\newcommand{\bmu}{{\bf \mu}}
\newcommand{\bpi}{{\bf \pi}}
\newcommand{\be}{\begin{equation}}
\newcommand{\ee}{\end{equation}}
\newcommand{\ben}{\begin{equation*}}
\newcommand{\een}{\end{equation*}}
\newcommand{\ba}{\begin{eqnarray}}
\newcommand{\ea}{\end{eqnarray}}
\newcommand{\ban}{\begin{eqnarray*}}
\newcommand{\ean}{\end{eqnarray*}}
\newcommand{\bay}{\begin{array}}
\newcommand{\eay}{\end{array}}
\newcommand{\bat}{\begin{tabular}}
\newcommand{\eat}{\end{tabular}}
\newcommand{\bi}{\begin{itemize}}
\newcommand{\ei}{\end{itemize}}
\newcommand{\bin}{\begin{enumerate}}
\newcommand{\ein}{\end{enumerate}}
\newcommand{\bfig}{\begin{figure}}
\newcommand{\efig}{\end{figure}}
\newcommand{\tr}{\text{tr}}
\newcommand{\Prob}{\text{Pr}}
\newcommand{\var}{\text{var}}
\newcommand{\impulse}{h(\tau,t)}
\newcommand{\intfy}{\int_{-\infty}^\infty}
\newcommand{\intT}{\int_{-T}^T}
\def\phase{e^ {-j2 {\pi} f_c {\tau} _k(t)}}
\def\carrier{e^ {j2 {\pi} f_ct} }
\def\vs2{\vspace*{2mm}}
\def\hs{\hspace {.2 in}}

\maketitle


\begin{abstract}
  We consider a \emph{cognitive network} consisting of $n$ random
  pairs of cognitive transmitters and receivers communicating
  simultaneously in the presence of multiple primary users. Of interest
  is how the maximum throughput achieved by the cognitive users scales
  with $n$. Furthermore, how far these users must be from a primary user
  to guarantee a given primary outage. Two scenarios are considered for
  the network scaling law: (i) when each cognitive transmitter uses
  constant power to communicate with a cognitive receiver at a bounded
  distance away, and (ii) when each cognitive transmitter scales its
  power according to the distance to a considered primary user, allowing
  the cognitive transmitter-receiver distances to grow. Using single-hop
  transmission, suitable for cognitive devices of opportunistic nature,
  we show that, in both scenarios, with path loss larger than 2, the
  cognitive network throughput scales {\em linearly} with the number of
  cognitive users. We then explore the radius of a \emph{primary
  exclusive region} void of cognitive transmitters. We obtain bounds on
  this radius for a given primary outage constraint. These bounds can
  help in the design of a primary network with exclusive regions,
  outside of which cognitive users may transmit freely. Our results show
  that opportunistic secondary spectrum access using single-hop
  transmission is promising.
\end{abstract}






\section{Introduction}
\label{sec:in}

The scaling laws of the capacity of ad-hoc wireless networks has been
an active area of research. Initiated by the work of Gupta and Kumar
\cite{Gupta_Kumar_00}, this area of research has been pursued  under a
variety of wireless channel models and communication protocol
assumptions \cite{Grossg_02, xie_kumar_04, xie_kumar_06, Kulkarni_04, 
   Agarwal_Kumar_04, Leveque_05, Jovicic_06, jovicic_scaling_06,
   Xue_Kumar_06, Frances_07, Ozgur_scaling_07, Ozgur_Leveque_Tse_07}. 
These papers usually assume $n$ pairs of homogeneous devices, thrown
at random in a plane, wishing to communicate. Each transmitter has a
single receiver, which may be located anywhere in the network. The
underlying question is how the total network throughput  (also called
the sum rate), or equivalently the per-user throughput, scales as the
number of communication pairs $n\rightarrow\infty$. This is
accomplished by either letting the density of nodes stay fixed and the
area increase with $n$ (extended network), or by fixing the network
area and letting the density increase with $n$ (dense
network). As transmitter-receiver pairs are selected at random in the
network, a  packet may require  transmission over multiple hops to
reach its destination. 

The throughput scaling in ad hoc networks depends greatly on the node 
distribution and the physical-layer processing capability, more
specifically the ability to cooperate among nodes. In the
interference-limited regime, in which no cooperation is allowed  
(except simple decode-and-forward), and all nodes treat other signals
as interference, the per-node throughput scales at most as $1/\sqrt{n}$ 
\cite{Gupta_Kumar_00}. If the nodes are uniformly distributed, a
simple nearest-neighbor forwarding scheme achieves a $1/(n\log(n))$ 
per-node throughput \cite{Gupta_Kumar_00}. When the nodes are
distributed according to a Poisson point process, a backbone-based
routing scheme achieves the per-node scaling of $1/\sqrt{n}$
\cite{Frances_07}, meeting the upper bound. On the other hand, when
nodes are able to cooperate, a much different  scaling law
emerges. Specifically, a hierarchical scheme can achieve a {\em
  linear} grow in the sum rate, corresponding to a constant per-node
throughput \cite{Ozgur_Leveque_Tse_07}. A key step in this scheme is
MIMO cooperation among nodes, which requires joint encoding and
decoding. The development of these scaling laws show that the
assumptions about the network and the nodes' signal processing
capability are crucial to the scaling law.

In this paper, instead of considering a homogeneous ad hoc wireless
network, we study a cognitive network consisting of two types of
users: primary and cognitive. Recent introduction of secondary
spectrum licensing 
necessitates the study of such cognitive networks. The cognitive users
opportunistically access the now-exclusive but under-utilized spectrum
of the primary users, while ensuring that any performance degradation
to the primary users is within an acceptable level. Other scenarios in
which two networks operate concurrently are also applicable.

Consider for example a TV station broadcasting in a now-exclusive,
licensed band. This band is wasted in geographic locations barely
covered by the TV signal. This prompts questions such as: can we allow 
other devices (cognitive users) to transmit in the same band as the TV
(primary users), provided their interference to any TV receiver is at
``an acceptable level''? If so, what is the minimum distance from the
TV station at which these devices can start transmitting? What are the
maximum rates that these devices can achieve by transmitting in the
TV band?

We formulate this problem from an information theoretic viewpoint for
a network with multiple primary and multiple cognitive users. We 
define the ``acceptable interference level'' to be a threshold on the
probability that the received signal (or rate) of a primary user is
below a certain level, provided that the primary receiver (Rx) is
within a radius of interest from its transmitter (Tx). This is
analogous to the concept of outage capacity. The radius of interest
specifies the primary exclusive region (PER), which are
non-overlapping for different primary transmitters. These PERs are
void of cognitive transmitters, but they can contain cognitive
receivers. We consider an extended network in which the cognitive
users are uniformly distributed such that their density is a constant.

Because of the opportunistic nature of the cognitive users, we
consider a network and communication model different from the
previously mentioned ad hoc networks. We assume that each cognitive 
transmitter communicates with a receiver within a {\em bounded
  distance} $D_{\max}$, using  {\em single-hop}
transmission. Being different from multi-hop communication in ad hoc 
networks, single hop communication appears suitable for cognitive
devices which are mostly short-range. Our results, however, are not
limited to short-range communication. There can be other cognitive
devices (transmitters and receivers) in between a Tx-Rx
pair. (This is different from the local scenarios of ad hoc networks,
in which every node is talking to its neighbor.) If the transmit power
model of the cognitive users is constant, then the maximum Tx-Rx
distance $D_{\max}$ stays constant. (In practice, we may preset a
$D_{\max}$ based on a large network and use the same value for all
networks of smaller sizes.) If we allow the cognitive devices to scale
its power according to the distance to the primary user, then
$D_{\max}$ may scale with the network size by a feasible
exponent. Other assumptions include a protected band around each 
receiver (primary or cognitive) to ensure that any interfering
transmitter is not at the same point as the interfered
receiver. Assuming no cooperation, the cognitive receivers simply
treats other users' signals as interference.

Within such a network, for both cases of constant and varying
cognitive transmit power, we find that the cognitive users' throughput
scales \emph{linearly} in the number of users $n$. Equivalently, as
$n\to\infty$, the {per user capacity} remains constant. Our
results thus indicate that an initial approach to building a scalable
cognitive network should involve limiting cognitive transmissions to a
single hop. This scheme appears reasonable for secondary spectrum
usage, which is opportunistic in nature.

The impact of cognitive users on the primary user is captured in the
expected amount of interference from the cognitive users. We derive
upper and lower bounds on this interference and show that the average
interference remains bounded irrespective of the number of cognitive
users. Based on this interference bounds, we provide an upper bound on 
the radius of the PER that satisfies the outage constraint on the
primary user's rate. The bound also allows us to study the
interdependence and trade-offs between the PER radius, the protected
band around each primary receiver and the primary transmit power.

The paper structure is as follows. In Section \ref{sec:pf}, we
introduce our network model and formulate the problem. In Section 
\ref{sec:sl}, we study the throughput scaling of the cognitive users
with constant transmit power in the presence of multiple primary
users. In Section \ref{sec:pe}, we examine the outage constraint on a
single primary user and derive an upper bound on  the radius of the
\emph{primary exclusive region}. In Section \ref{sec:ps}, we
investigate the option of allowing the cognitive users to scale the
transmit power according to the distance to the primary user. In
Section \ref{sec:cn} we make our conclusions.


\section{Problem Formulation}
\label{sec:pf}

Consider a cognitive network with two types of users: primary 
and cognitive users. We address two main questions. First, how the
total throughput of the cognitive users scales with network size,
given the presence of the primary users. Second, at what distance from 
the primary users can these cognitive users operate to ensure a
maximum outage probability for the primary user. The users are not
allowed to cooperate, hence the network is interference-limited. We
will first discuss the network model and the channel and signal
models. We then formulate specifically each of the two criteria:
cognitive users throughput, and the allowable distance from the
primary user.


\subsection{Network model}

We consider an extended network with all transmitters and receivers
located on a plane. With fixed nodes densities, the network size grows 
with the number of nodes. As a specific instance, we consider a
circular network with radius $R$. To scale the number of cognitive and
primary users, we let $R$ increase. Other shapes also produce a
similar scaling law.

\begin{figure}[tb]
\centerline{\epsfig{figure=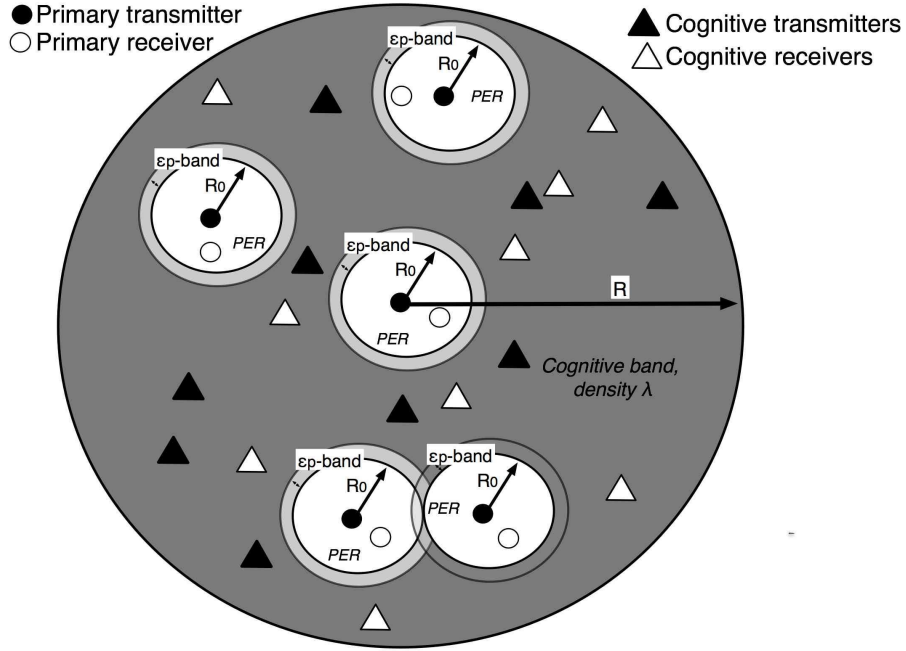 ,width=12cm}}
  \caption{A cognitive network consists of multiple primary users and
  multiple cognitive users. Each primary transmitter Tx$_p^i$ is at the
  center of a \emph{primary exclusive region} (PER) with radius $R_0$,
  which contains its intended receiver. These PERs are
  non-overlapping. Surrounding each PER is a protected band of width
  $\epsilon_p>0$. Outside the PERs and the protected bands, $n$
  cognitive transmitters are distributed randomly and uniformly with
  density $\lambda$. The placement model for the cognitive users is
  illustrated in Figure \ref{fig:cog_example}.}
  \label{fig:overall_net}
\end{figure}

We introduce our network model in Figure \ref{fig:overall_net}.
Within the network, there are $m$ primary users and $n$ cognitive
users. Let Tx$_p^i$ and Rx$_p^i$ denote a primary transmitter and its
intended receiver ($i=1,2,\cdots, m$), and Tx$_c^j$ and Rx$_c^j$ for  
the cognitive transmitter and receiver ($j=1,2,\cdots, n$). Each
primary transmitter is located at the center of a primary exclusive
region (PER) with radius $R_0$; the corresponding primary receiver can
be anywhere within this region. This model is based on the premises
that the primary receiver location may not be known to the cognitive
users. Such a setup is typical in broadcast scenarios, such as often
found in cellular or TV networks. Hence we choose to center the PER
circle on the primary transmitter Tx$_p^i$ (for example, a base
station) rather than the receiver Rx$_p^i$. These PERs of radius $R_0$
are non-overlapping. WOLG, we also assume that there is a PER at the
center of the network, with the transmitter Tx$_p^1$ and receiver
Rx$_p^1$. Other than that, we make no specific assumptions about the 
placement of the PERs, or the primary transmitters. This means their
locations are arbitrary.

Around each primary receiver we assume there is a circle of radius
$\epsilon_p>0$ in which no interfering cognitive transmitter may
lie. (The cognitive receivers, however, can lie within this
$\epsilon_p$ circle.) Because the location of the primary receiver is 
unknown to the cognitive transmitters, this assumption results in a 
protected band of width $\epsilon_p$ around each PER, inside which no 
cognitive transmitters may operate. We will later design this radius
$\epsilon_p$ in conjunction with $R_0$ and other system parameters to
meet certain primary outage constraints. 

Similarly, we assume that all cognitive receivers have a protected
circle of radius $\epsilon_c>0$ around them, in which no interfering,
either primary or cognitive, transmitters may lie ($\epsilon_c$ may be
different from $\epsilon_p$). These practical constraints simply
ensure that the interfering transmitters and receivers are not located
at exactly the same point.

\begin{figure}[tb]
\centerline{\epsfig{figure=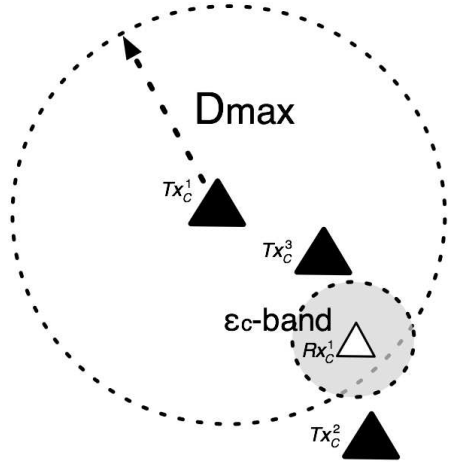,width=6cm}}
  \caption{Cognitive user model: Each cognitive transmitter
    Tx$_c^i$ wishes to transmit to a single cognitive receiver
    Rx$_c^i$, which lies within a distance $\leq D_{max}$ away. Each
    cognitive receiver has a protected circle of radius
    $\epsilon_c>0$, in which no interfering transmitter may operate.}
  \label{fig:cog_example}
\end{figure}

All cognitive transmitters are distributed outside the PERs encircled
by an $\epsilon_p$-band. We assume that the cognitive transmitters are 
randomly and uniformly distributed with constant density
$\lambda$. The cognitive receivers, however, can be anywhere in the
network (subject to the $\epsilon_c$ protected distance), including
inside the PERs. We assume that each cognitive receiver is within a
$D_{\max}$ distance from its transmitter. Depending on the transmit
power of the cognitive users, $D_{\max}$ may scale with the network
size (as analyzed later). Figure \ref{fig:cog_example} provides an
example of a such cognitive Tx-Rx layout.

Table \ref{table:variables} summarizes the network notation.
\begin{table}
\begin{center}
\begin{tabular}{|ll|}
\hline
  Primary user $i$th transmitter and receiver & Tx$_p^i$, Rx$_p^i$\\
  Cognitive user $i$th transmitter and receiver & Tx$_c^i$, Rx$_c^i$\\
  Outer radius for cognitive transmission & $R$\\
  Number of primary users & m\\ 
  Number of cognitive users & n\\ 
  Primary exclusive region radius & $R_0$\\
  Maximum cognitive Tx$_c^i$-Rx$_c^i$ distance & $D_{\max}$\\
  Minimum Tx$_c^i$-Rx$_p^k$ distance & $\epsilon_p$\\
  Minimum Tx$_c^i$-Rx$_c^k$ distance ($i\neq k$) & $\epsilon_c$\\
  Cognitive user density & $\lambda$\\
  \hline
\end{tabular}
\caption{Network notation.}
\label{table:variables}
\end{center}
\end{table}

The introduced network is a general model with multiple primary and 
multiple cognitive users, which we will use to study the scaling law
of the cognitive users. We then use a special case of this model --
with only a single primary user at the center -- to study the radius
of the primary exclusive region (the minimum distance from a primary
user at which a cognitive user can operate).


\subsection{Channel and signal models}
We consider a path-loss only model for the wireless channel. Given a
distance $d$ between the transmitter and the receiver, the channel $h$ 
is therefore given as
\be
  h = \frac{A}{d^{\alpha/2}} \label{eq:pathloss}
\ee
where $A$ is a frequency-dependent constant and $\alpha$ is the power
path loss. In subsequent analysis, we normalize $A$ to be 1 for
simplicity. We consider $\alpha > 2$ which is typical in practical 
scenarios.

We assume that the channels between different transmitters and
receivers are independent. Furthermore, they all undergo independent
zero-mean additive white Gaussian noise of power $\sigma^2$. We define
the notation for selected channels in Table \ref{table:channels}.
\begin{table}[h]{
\begin{center}
\begin{tabular}{|ll|}
\hline
  Channel from cognitive Tx$_c^i$ to cognitive Rx$_c^j$ & $h_{ij}$\\
  Channel from primary Tx$_p^i$ to cognitive Rx$_c^j$ & $g_{ij}$\\
  Channel from the center primary Tx$_p^1$ to its primary Rx$_p^1$ & $h_0$\\
  Channel from cognitive Tx$_c^i$ to primary Rx$_p^1$ & $g_i$\\
\hline
\end{tabular}
\caption{Channel notation.}
\label{table:channels}
\end{center}}
\end{table}

For the signal model, we assume no multiuser detection. Thus each
user, either primary or cognitive, has no knowledge of other users'
signals and treats their interference as noise. We assume that each
primary user signal is constrained by a constant power $P_0$, and each
cognitive user by $P$. (Later on we will consider two cases: $P$ is
constant, and $P$ is variable with distance). Furthermore, the signals
of different users are statistically independent. With a large number
of users, independent and power-constrained, their interference to a
receiver will be (approximately) Gaussian. Thus the optimal transmit
signals for both types of users are zero-mean Gaussian.


\subsection{The cognitive network throughput}
Consider the transmission rate of a cognitive user in the presence of
other cognitive users and multiple primary users. Denote $I_{ci}$ 
and $I_{pi}$ ($i = 0, \ldots, n$) as the total interference power to
cognitive user $i$ from other cognitive transmitters and from the
primary transmitters, respectively. 
Based on Table \ref{table:channels}, these interference powers can be
written as
\ba
  I_{ci} = \sum_{j\neq i} P |h^c_{ji}|^2 \;, \quad
  I_{pi} = \sum_{j=1}^n P_0 |g^c_{ji}|^2 . \label{cog_interf}
\ea
With Gaussian signaling and transmit power $P$, the rate of cognitive
user $i$ can then be written as
\be
  C_i = \log\left( 1 + \frac{P |h^c_{ii}|^2}{I_{ci} + I_{pi} 
  + \sigma^2} \right) \;, \quad i = 1\ldots n.
  \label{cog_rate}
\ee
Because of the random placement of cognitive users, $I_{ci}$ is a
random variable. The rate $C_i$ therefore is also random.

Define the {\em average} total throughput of the cognitive users as 
\be
  S_n = \sum_{i=1}^n E[C_i].
  \label{eq:sum_rate}
\ee
An equivalent measure is the per-user throughput defined as
\be
  T_n = \frac{1}{n} S_n.
  \label{eq:per_user_rate}
\ee
We are interested in how the average sum rate (\ref{eq:sum_rate}),
equivalently the per-user throughput (\ref{eq:per_user_rate}), scales
as $n\to\infty$.


\subsection{The primary exclusive region}
To study the radius of the primary exclusive region, we consider a
special case of the network with only a single PER at the center. In
other words, we consider only Tx$_p^1$ at the center of the network
and its receiver Rx$_p^1$ within a radius $R_0$ from the primary 
transmitter. The main reason is that we focus on the impact on a
primary user of the addition of cognitive users. Without these
cognitive users, the primary network would operate with noise and the
usual interference from the other primary users. Hence this special
case can also be thought of as approximating the noise power to
include the interference from other primary users to the considered
user.

The radius $R_0$ of the primary exclusive region is determined by the
outage constraint on the primary user given as 
\ben
  \text{Pr}\left[ \text{primary user's rate} \leq C_0 \right] \leq \beta
\een
where $C_0$ and $\beta$ are pre-chosen constants. This constraint
guarantees the primary user a rate of at least $C_0$ for all but
$\beta$ fraction of the time.

Denote $h_0$ as the channel of the considered primary user, and $g_i$
as the channel from cognitive transmitter $i$ to this user's
receiver (as in Table \ref{table:channels}). The interference power
from the cognitive users to the considered primary user is
\be
  I_0 = \sum_{i=1}^n P |g_i|^2
  \label{cog_prime_interf}
\ee
Again this interference power is random because of the random
placement of the cognitive users. With Gaussian signaling, the rate of
this primary user can be written as
\ben
  C_p = \log\left( 1 + \frac{P_0 |h_0|^2}{I_0 + \sigma^2} \right) \;.
\een
This rate is random because of random interference $I_0$. The outage
constraint can now be written as
\be
  \text{Pr}\left[ \log\left( 1 + \frac{P_0 |h_0|^2 }{I_0 + \sigma^2 }
    \right) \leq C_0 \right] \leq \beta .
  \label{outage_constr}
\ee
Since we consider channels with only path loss, outages occur here are
not because of fading as in traditional schemes, but because of the
random placement of cognitive users.


\subsection{Single-primary network with cognitive power scaling}
When considering a network with a single primary user at the center, a
feasible option is to allow the cognitive transmitters to scale their 
power according to the distance from the primary user. Specifically,
the transmit power $P$ of a cognitive user is now a function of the
radius $r$, at which this cognitive user is located, as
\be
  P(r) = P_c r^\gamma
  \label{cog_power_scaling}
\ee
for some constant power $P_c$ and a feasible power exponent
$\gamma$ (which will be analyzed later). Similar to the case of
constant cognitive transmit power, we will also examine the throughput 
scaling of the cognitive users and the primary exclusive radius in
this case (albeit both objectives with a single-primary network model).


\section{The Scaling Law of a Cognitive Network with constant power}
\label{sec:sl}
In this section, we study the throughput of the cognitive users with
constant transmit power, assuming multiple primary users. In
particular, we examine the throughput scaling law as the number of
cognitive users $n$ increases to infinity. We first establish upper
and lower bounds to the per-user throughput and then show that both
bounds scale with the same order, which then becomes the scaling order
of the throughput itself.

\subsection{Lower bound on the cognitive per-user capacity}
To derive a lower bound on the capacity of a cognitive user, we study
an upper bound on the interference to a cognitive receiver. This
includes the interference from the primary users and from the
cognitive users.

\subsubsection{Interference from the primary users}
\label{sec:cog_interf_primary}
We assume that the primary users must be spaced such that the primary 
exclusive regions (with radius $R_0$) are non-overlapping. Two PERs
closest to each other may have the boundaries (circles) touching at
one point, as an example shown in Figure \ref{fig:overall_net}. We
shall examine the densest placement of the primary users to upper
bound the interference from them to a cognitive receiver.

Consider boundary circles of the PER with radius $R_0$. The tightest
circle packing is according to the hexagon lattice \cite{Conway_92}, as
shown in Figure \ref{fig:hex_lattice}, in which the three bold circles
represents the PERs with radius $R_0$. Since each cognitive receiver
has a protected radius of $\epsilon_c$, the worst case cognitive
receiver will be on a circle of radius $\epsilon_c$ around a primary
transmitter. We are interested in the interference from all the
primary transmitters, located on the hexagon lattice, to this
cognitive receiver.
\bfig[tb]
\centerline{\epsfig{figure=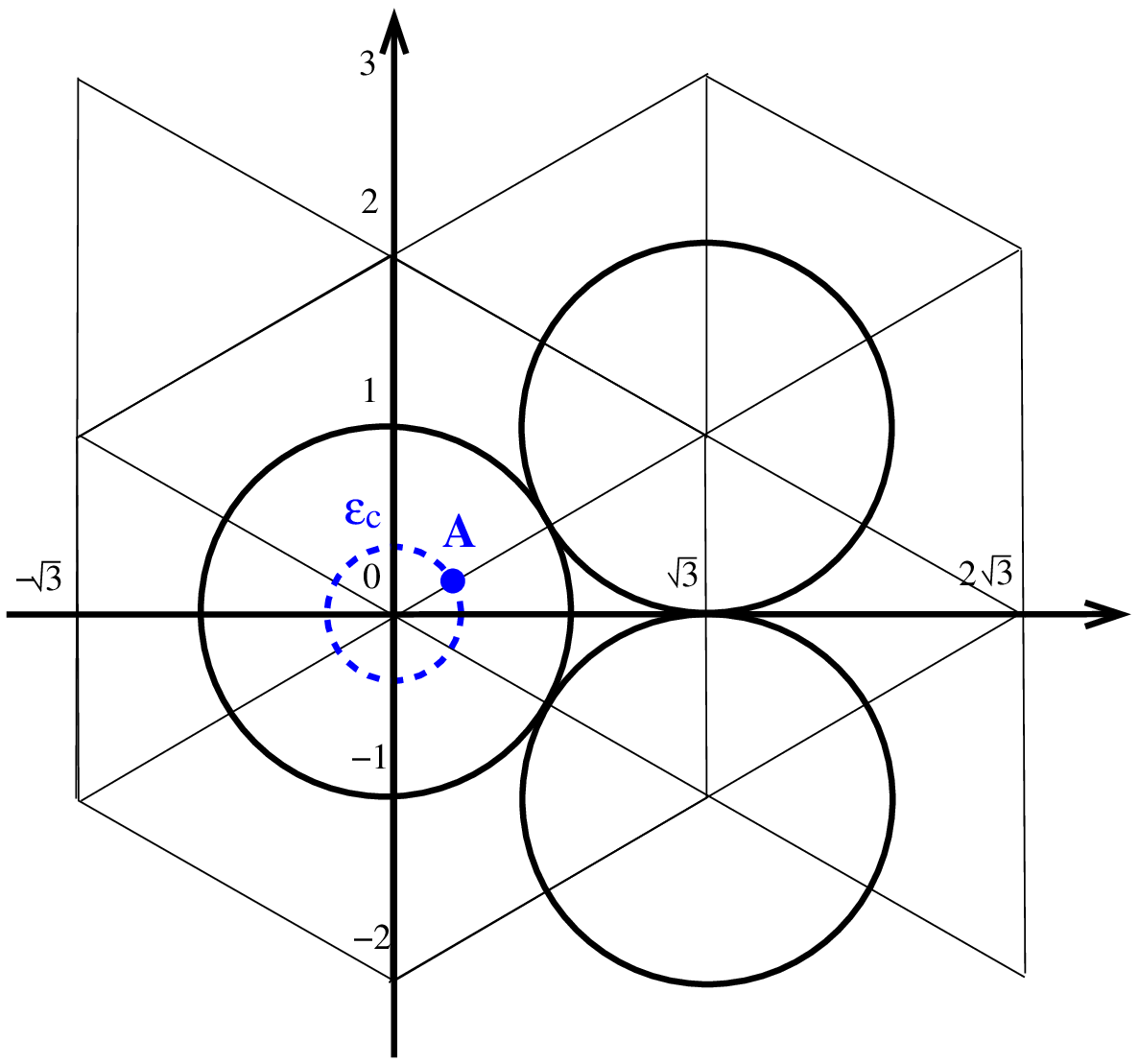 ,width=8.6cm}}
  \caption{Worst-case interference when the primary users locate on a 
    hexagon lattice.}
  \label{fig:hex_lattice}
\efig

To calculate the interference, we superimpose x-y axes on the hexagon 
topology with the origin at a primary transmitter and with normalized
length (1 unit length = $R_0$), as shown in Figure
\ref{fig:hex_lattice}. Consider the circle of radius $\epsilon_c$
around the origin. (For practical reasons, we consider $\epsilon_c < 
R_0$.) Let A be a point on this circle at an angle $\theta$ to the
x-axis. The interference from all primary transmitters to A is 
\ba
  I_A &=& \sum_{m=-\infty}^\infty \sum_{k=-\infty}^\infty
  \frac{1}{\left[\left(2\sqrt{3}k-\epsilon_c\cos\theta\right)^2 +
      \left(2m-\epsilon_c\sin\theta\right)^2\right]^{\alpha/2}} \nonumber \\
   &+& \sum_{m=-\infty}^\infty \sum_{k=-\infty}^\infty 
    \frac{1}{\left[\left(\sqrt{3}(2k+1)-\epsilon_c\cos\theta\right)^2 +
	\left(2m+1-\epsilon_c\sin\theta\right)^2\right]^{\alpha/2}} .
    \label{I_A}
\ea
For any $\theta \in [0,2\pi]$, each of the above summations is bounded
for $\alpha>2$, as shown in the Appendix. Let $I_P$ be the worst-case 
interference from the primary users, 
\be
  I_P = \max_\theta I_A
  \label{I_P}
\ee
then $I_p$ is also bounded. Thus the total interference from all
primary users to any cognitive receiver is bounded.

\subsubsection{Worst-case interference from cognitive users}
An upper bound is obtained by filling all primary exclusive regions
with cognitive transmitters. 
Since the cognitive user density is constant, this filling increases
the number of cognitive users at most by a scaling factor (the ratio
between the area of the PERs and the area occupied by cognitive
users).

Now consider a uniform network of $n$ cognitive users. The worst case
interference would then be to a cognitive receiver at the center of
the network (without loss of generality assumed to be Rx$_c^1$). From
the considered receiver, draw a circle of radius $R$ that covers all
other cognitive transmitters. With constant user density of $\lambda$
users per unit area, then $R^2$ grows linearly with $n$ (in other
words, $R^2$ is of order $n$).

To see that this case is indeed the worst interference from cognitive
users, consider another cognitive receiver (Rx$_c^2$) that is not at
the center of the network. Again draw a circle of radius $R$ centered at
Rx$_c^2$. Since this receiver is not at the center of the network, the
circle will not cover all cognitive transmitters. The interference to
Rx$_c^2$ is then increased by moving all the transmitters from outside
this new circle (area A in Figure \ref{interf_cicles}) to inside the
circle (area B in Figure \ref{interf_cicles}), resulting in the same
interference as to Rx$_c^1$.

\bfig[tb]
\centerline{\epsfig{figure=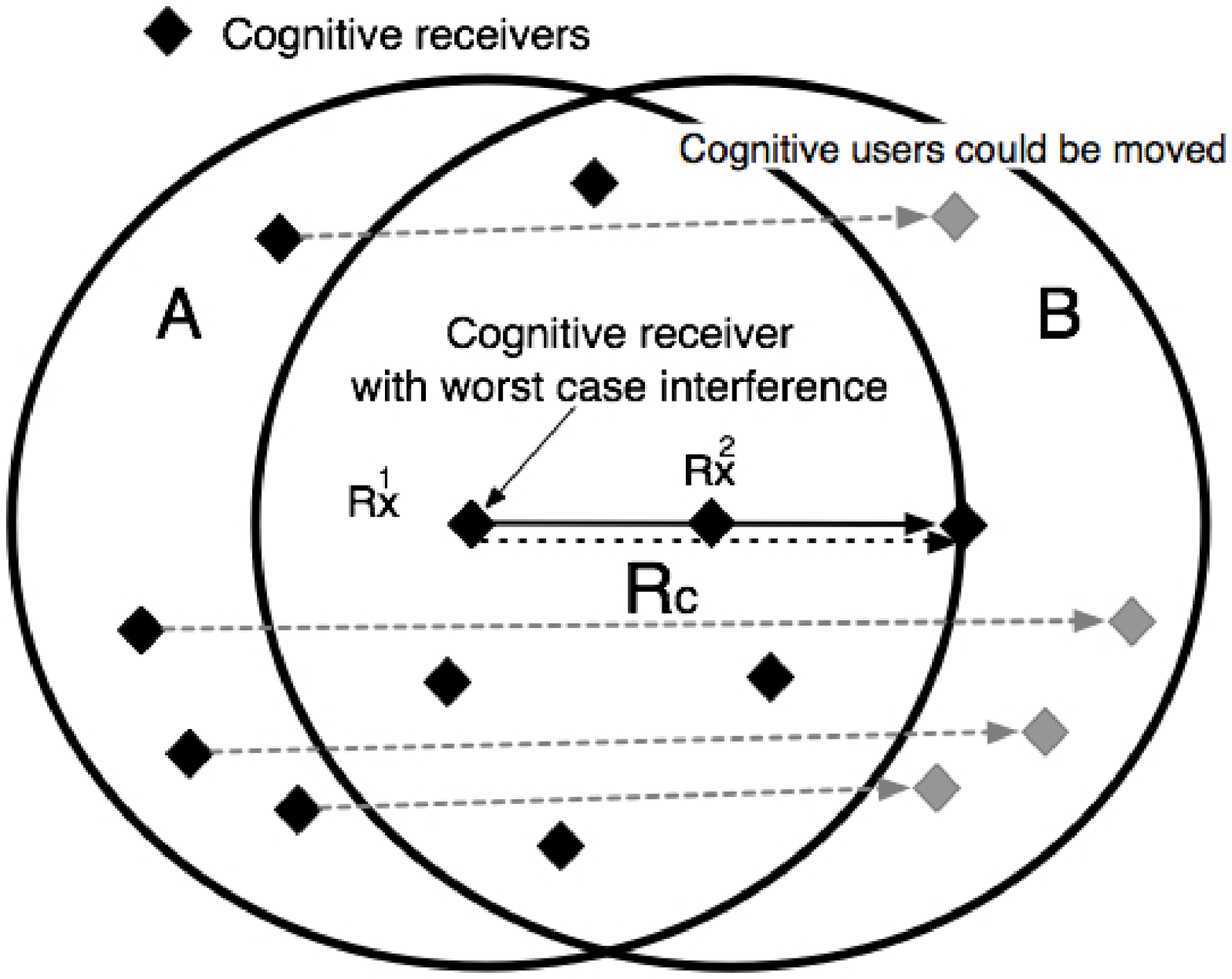 ,width=7.5cm}}
  \caption{Worst-case interference to a cognitive receiver.}
  \label{interf_cicles}
\efig

Consider an interfering cognitive transmitter located randomly within
the circle of radius $R$ from the considered receiver. With uniform
distribution, the distance $r$ between this interfering transmitter
and this receiver has the density
\be
  f_r(r) = \frac{2r}{R^2-\epsilon_c^2} \;, \quad \epsilon_c \leq r \leq
  R \:.
  \label{r_density}
\ee
The {\em average interference} from this transmitter to the considered 
receiver therefore is
\begin{align}
  I_{\text{avg},1} 
  & = \int_{\epsilon_c}^{R} \frac{2rP}{(R^2-\epsilon_c^2) r^\alpha} dr
  \nonumber \\
  &= \frac{2P}{(R^2-\epsilon_c^2)(\alpha-2)} \left(\frac{1}{\epsilon_c^{\alpha-2}} - 
    \frac{1}{R^{\alpha-2}} \right) . 
    \label{avg_cog_interf_1}
\end{align}
The average total interference from all other cognitive transmitters
to the considered receiver then becomes 
\[  I_{\text{avg},n} = n I_{\text{avg},1}\:. \]
But $\lambda\pi (R^2-\epsilon_c^2) = n$, thus
\be
  I_{\text{avg},n} 
  = \frac{2\pi\lambda P}{(\alpha-2)} 
  \left(\frac{1}{\epsilon_c^{\alpha-2}} - \frac{1}{R^{\alpha-2}} \right) .
  \label{avg_cog_interf_n}
\ee
For any cognitive receiver, its average interference is upper-bounded
by $I_{\text{avg},n}$, that is
\be
  E[I_i] \leq I_{\text{avg},n} \:.
\ee
As $n\to\infty$, provided that $\alpha > 2$, this average interference
to the cognitive receiver at the center approaches a constant as
\be
  I_{\text{avg},n} 
  \; \stackrel{n\to\infty}{\longrightarrow} \;
  \frac{2\pi\lambda P}{(\alpha-2)\epsilon_c^{\alpha-2}}
  \stackrel{\triangle}{=} I_{\infty} .
  \label{I_infty}
\ee

\subsubsection{Lower bound on the cognitive per-node throughput}
Now consider the rate of the $i$th cognitive user given in (\ref{cog_rate}).
Since the distance between a cognitive transmitter and its intended
receiver is bounded by $D_{\max}$, we have $|h_{ii}|^2 \geq
1/D_{\max}^\alpha$. Denote the minimum cognitive received power as
$P_{r,\min} = P/D_{\max}^\alpha$. Given that the interference from
the primary users is bounded by $I_P$ in (\ref{I_P}), then 
\be
  C_i \geq \log\left( 1 + \frac{P_{r,\min}}{\sigma^2 + I_P + I_i} \right).
\ee
Noting that $\log(1 + a/x)$ is convex in $x$ for $a > 0$, by Jensen's
inequality, we have
\ben
  E\log\left( 1 + \frac{a}{X} \right) \geq \log\left(1 + \frac{a}{EX}
  \right) .
\een
Thus the average rate of each cognitive user satisfies
\ba
  E[C_i] &\geq& \log\left( 1 + \frac{P_{r,\min}}{\sigma^2 + I_P + 
    E[I_i]} \right) \nonumber \\
  &=& \log\left( 1 + \frac{P_{r,\min}}{\sigma^2 + I_P + 
    I_{\text{avg},n}} \right).
\ea
As $n\to\infty$, the lower bound approaches a constant as
\be
  E[C_i] \geq \log\left( 1 + \frac{P_{r,\min}}{\sigma^2 + I_P + 
    I_{\infty}} \right) \stackrel{\triangle}{=} \bar{C_1}
  \label{rate_lower_bnd}
\ee
where $I_{\infty}$ is defined in (\ref{I_infty}). Thus the average
per-user rate of a cognitive network remains at least a constant as
the number of users increases.

\subsection{Upper bound on the network sum capacity}
A trivial upper-bound can be obtained by removing the interference
from all other cognitive users. Assuming that the capacity of a single
cognitive user under noise alone is bounded by a constant, then the
total network capacity grows at most linearly with the number of users.

\subsection{Linear scaling law of the cognitive network average throughput}
From the above lower and upper bounds, we conclude that the average
sum throughput of the cognitive network grows linearly in the number
of users
\be
  E[S_n] = n K \bar{C_1}
  \label{avg_rate_scaling}
\ee
for some constant $K$, where $\bar{C_1}$ defined in (\ref{rate_lower_bnd})
is the achievable average rate of a single cognitive user under
constant noise and interference power. In other words, the average
per-user rate stays constant as the number of users increases.

\subsection{The concentration of the network throughput around its mean}
\label{sec:concent}
Given that the {\em average} cognitive users' throughput scales
linearly with the number of users, the concentration of the throughput
around its mean is also of interest. This concentration provides the 
probability that the throughput of a specific network (with a
realization of the cognitive user locations) scales at the same rate
as the mean throughput. 
Suppose this specific throughput can be written as $S_n = E[S_n] 
+ \Delta = n K \bar{C}_1 + \Delta$ for some real $\Delta$. Then we
need to show that with high probability, $\frac{|\Delta|}{n}$ approach
0 as $n\to\infty$.

Specifically, for a $\delta > 0$, we examine
\ben
  P_\delta \stackrel{\triangle}{=} \Pr\left[ \frac{1}{n}|S_n - E[S_n]|
    \geq \delta \right],
\een
where $S_n = \sum C_i$ as given in (\ref{eq:sum_rate}). Since we
consider only channels with path loss, the rate $C_i$ of each
cognitive user in (\ref{cog_rate}) is bounded. Thus $C_i$ is a random
variable with finite mean and finite variance. Furthermore, the $C_i$
are i.i.d. Thus by the central limit theorem, as $n\to\infty$, the sum 
throughput (\ref{eq:sum_rate}) can be approximated as a Gaussian
random variable. Then the following inequalities hold:
\ban
  P_\delta &\preceq& \frac{1}{\sqrt{2\pi}} \int_\frac{n\delta}{\sqrt{\var(S_n)}}^\infty  e^{-z^2/2} dz \\
    &\leq& \frac{\sqrt{\var(S_n)}}{n\delta\sqrt{2\pi}} \int_\frac{n^2\delta^2}{{2\var(S_n)}}^\infty
    e^{-w} dw \\    
   &=& \frac{\sqrt{\var(S_n)}}{n\delta \sqrt{2\pi}} \exp\left(-
    \frac{n^2\delta^2}{2\var(S_n)} \right) , 
\ean
where the notation $\preceq$ indicates that the inequality holds  in
the limit as $n\to\infty$. Now suppose $K_2 > 0$ is an upper bound on
the variance of $C_i$, that is $\var(C_1) \leq K_2$, then $\var(S_n) =
n\var(C_i) \leq nK_2$. Therefore
\ban
  \frac{\sqrt{\var(S_n)}}{n\delta \sqrt{2\pi}} \exp\left(-
  \frac{n^2\delta^2}{2\var(S_n)} \right) \leq 
  \frac{\sqrt{nK_2}}{n\delta \sqrt{2\pi}} \exp\left(-
  \frac{n^2\delta^2}{2nK_2} \right) 
  \stackrel{n\to\infty}{\longrightarrow} 0 
\ean
This means that any deviation of the throughput of a specific network
from its mean scales sub-linearly. Thus with high probability, the
total throughput of the cognitive users in a specific network scales
linearly with the number of users.


\section{The Primary Exclusive Region}
\label{sec:pe}

\begin{figure}[tb]
\centering{\epsfig{figure=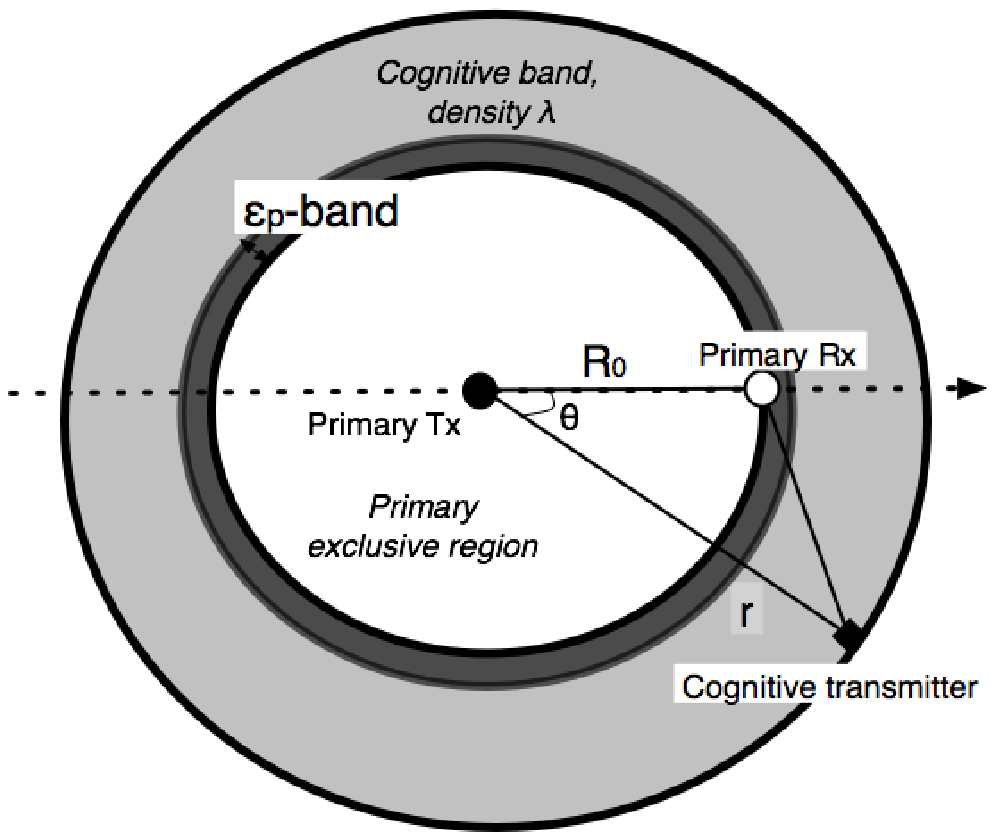,width=7cm}}
  \caption{Worst-case interference to a primary receiver: the receiver
  is on the boundary of the primary exclusive region of radius
  $R_0$. We seek to find $R_0$ to satisfy the outage constraint on the
  primary user.}
  \label{cognet_prime_interf}
\end{figure}

We next study the relation between the primary exclusive region radius
$R_0$ and the primary receiver guard band width $\epsilon_p$. For this
part, we focus on a single primary transmitter Tx$_p^1$ at the center
of the network of radius $R$. This primary transmitter is surrounded
by a single PER and an $\epsilon_p$-width transmission-free band, as
shown in Figure \ref{cognet_prime_interf}. Outside this PER are the
cognitive users. The cognitive transmitters must lie outside the
circle of radius $R_0+\epsilon_p$. In other words, these transmitter
cannot be placed in the transmission-free $\epsilon_p$-band, which is
a valid assumption in all scenarios in which the cognitive transmitter
is forbidden to be placed in exactly the same location as the primary
receiver. (Ideally, there needs to be only a protected circle of
radius $\epsilon_p$ around the primary receiver. But since we assume
that the cognitive users may not know the location of this primary
receiver, we impose a whole transmission-free band.) A cognitive
receiver, however, can lie in the $\epsilon_p$ band. To study the
primary exclusive region, we consider the worst case scenario in which
the primary receiver Rx$_p^1$ is at the edge of this region, on the
circle of radius $R_0$, as shown in Figure \ref{cognet_prime_interf}. The
outage constraint must also hold in this (worst) case, and we find a
bound on $R_0$ to ensure this.

Consider interference at the primary receiver on the boundary of the
PER from a cognitive transmitter at radius $r$ and angle $\theta$. The
distance $d(r,\theta)$ (the distance depends on $r$ and $\theta$)
between this interfering transmitter and the primary receiver satisfies
\ben
  d(r,\theta)^2 = r^2 + R_0^2  - 2R_0 r \cos\theta \:.
\een
For uniformly distributed cognitive users, $\theta$ is uniform in $[0,
2\pi]$, and $r$ has the density
\ben
  f_r(r) = \frac{2r}{R^2-(R_0+\epsilon_p)^2} .
\een
The expected interference \emph{power} experienced by the primary
receiver from all $n=\lambda\pi(R^2-(R_0+\epsilon_p)^2)$ cognitive
users is then given as
\begin{align}
  E[I_0] &= n \int_{R_0+\epsilon_p}^{R} \int_0^{2\pi} \frac{P}{
    d(r,\theta)^{\alpha}} f_r(r) f_\theta(\theta) \; dr \; d\theta \nonumber \\
  &= \int_{R_0+\epsilon_p}^R  \int_0^{2\pi}
  \frac{\lambda r P \; dr\; d\theta}{(r^2 + R_0^2 - 2R_0 r
    \cos\theta)^{\alpha/2}}. \label{eq:int1}
\end{align}

For $\alpha=2k$ with integer $k$, we can calculate $E[I_0]$
analytically. As an example, for $\alpha=4$, we obtain the values of
$E[I_0]$ as
\begin{align}
E[I_0]_{\alpha=4} &  = \lambda\pi
P\left[-\frac{R^2}{(R^2-R_0^2)^2}+\frac{(R_0+\epsilon_p)^2}{\epsilon_p^2(2R_0+\epsilon_p)^2}
  \right]. \label{I0alpha4}
\end{align}
The derivation is in the Appendix. Letting $R\rightarrow \infty$, this
average interference becomes
\begin{align}
E[I_0]_{\alpha=4}^{\infty} & = \lambda\pi
P\left[\frac{(R_0+\epsilon_p)^2}{\epsilon_p^2(2R_0+\epsilon_p)^2}
  \right] 
 \label{I0alpha4infty}
\end{align}

Next, we derive bounds on this expected interference power $E[I_0]$ at
the primary receiver for a general $\alpha$. We use these bounds to
analyze the interference versus the radius $R_0$ and the path loss 
$\alpha$. We then relate the outage probability to the average
interference through the Markov inequality and establish an explicit
dependence of $R_0$ on $\epsilon_p$ and other design parameters.

\subsection{Upper and lower bounds on the average interference}

In this subsection we obtain two lower bounds and an upper bound on $E[I_0]$.

\subsubsection{A first lower bound on $E[I_0]$}

\begin{figure}
\centerline{\epsfig{figure=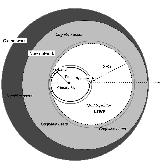, width=7cm}}
\caption{A lower bound on the expected interference at the primary Rx
  is obtained by forming a cognitive-free circle of radius $2R_0$
  around the primary receiver and reducing the network radius, now
  centered at the primary receiver, to $R-R_0$. All cognitive
  transmitters now lie within these two new boundaries.}
\label{fig:lower_bound1}
\end{figure}

A first lower bound on $E[I_0]$ can be established by re-centering the
network at the {\em primary receiver} Rx$_p^1$. We then make a new
exclusive region of radius $2R_0$, and a new outer radius of $R-R_0$,
both centered at Rx$_p^1$, as shown in Figure
\ref{fig:lower_bound1}. The set of cognitive users included in the 
new ring will be a subset of the original, making the interference a
lower bound as
\ba
  E[I_0]_\text{LB1} &=& \int_{2R_0+\epsilon_p}^{R-R_0} \frac{2\pi\lambda
    Pr}{r^\alpha}dr \nonumber \\
  &=& \frac{2\pi\lambda P}{\alpha-2} \left(\frac{1}{(2R_0+\epsilon_p)^{\alpha-2}} -
    \frac{1}{(R-R_0)^{\alpha-2}} \right). \label{I0_lower2}
\ea

As $R\to\infty$, this bounds approach the limit:
\ba
E[I_0]_{LB 1}^{\infty} &=& \frac{2\pi P\lambda}{\alpha-2}
  \frac{1}{(2R_0+\epsilon_p)^{\alpha-2}} \label{I0_lower1_Rinfty}
\ea

\subsubsection{A second lower bound on $E[I_0]$}

\begin{figure}
\centering{\epsfig{figure=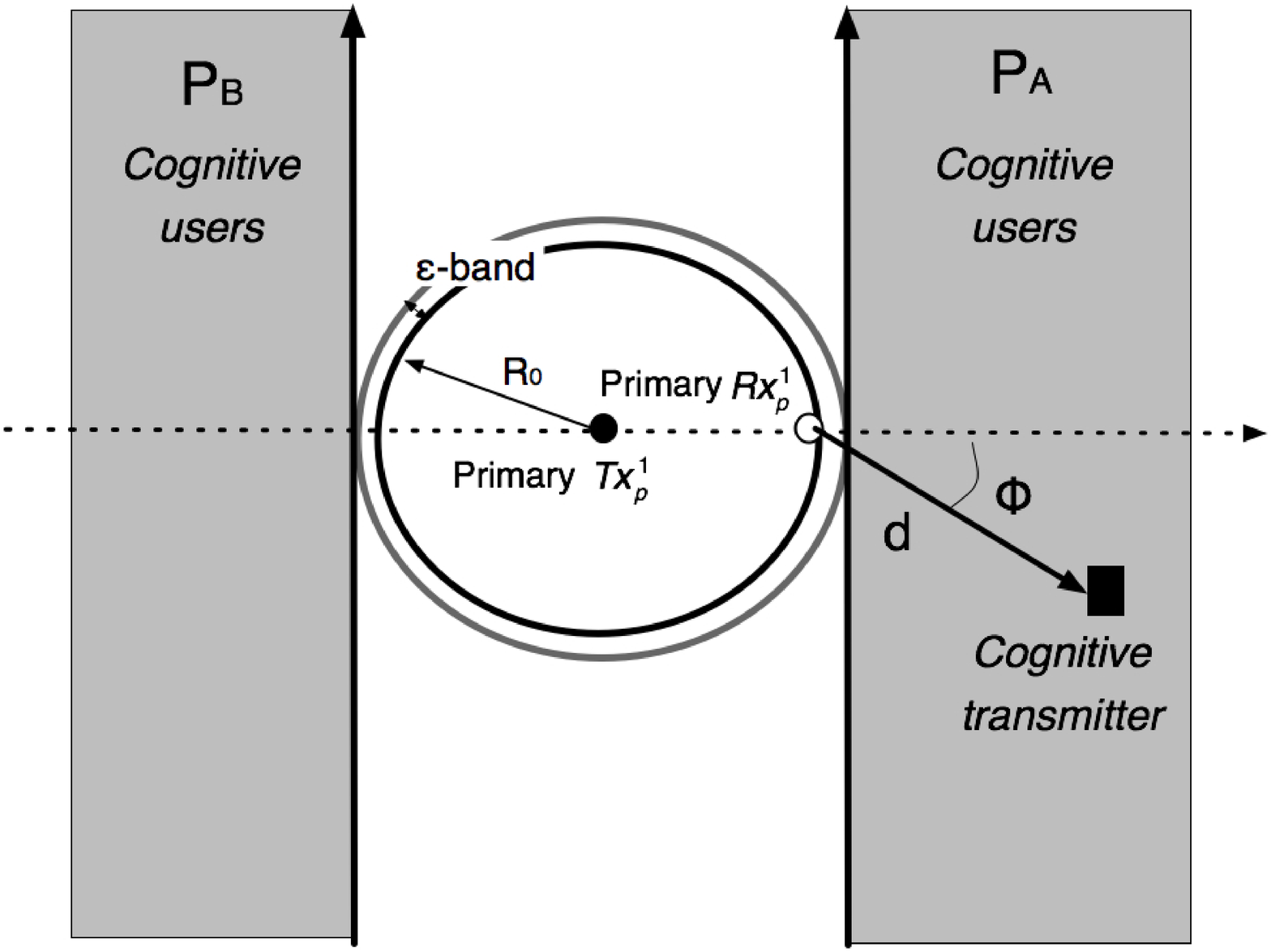, width=7cm}}
\caption{Another lower bound on the expected interference at the
  primary Rx  is obtained by  approximating the interference region by
  two half-planes $P_A$ and $P_B$. The region between these planes is
  free from cognitive transmitters.}
\label{fig:lower_bound2}
\end{figure}

Another lower bound on the interference can be derived by
approximating the interference region by two half-planes, similar to
\cite{Sahai_05}. As illustrated in Figure \ref{fig:lower_bound2},
consider only interference from the cognitive users in the two
half-planes $P_A$ and $P_B$ which touch the circle of radius
$R_0+\epsilon_p$. Consider a line in $P_A$ that makes an angle $\phi$
at Rx$_p^1$, the distance $d$ from any point on this line to Rx$_p^1$
satisfies $\frac{\epsilon_p}{\cos(\phi)} \leq d < \infty$. Since the
cognitive users are distributed uniformly, as $R\to\infty$, the
distribution of $d$ becomes similar to the distribution of $r$ given
in (\ref{r_density}), and $\phi$ will be uniform in $[-\frac{\pi}{2},
\frac{\pi}{2}]$. Similar analyses hold for $P_B$. Hence the average total 
interference from the cognitive users in $P_A$ and $P_B$ to Rx$_p^1$ is
\ba
  E[I_0]_\text{LB2} &=& P\lambda \left(\int_{-\frac{\pi}{2}}^{\frac{\pi}{2}}
    \int_{\frac{\epsilon_p}{\cos(\phi)}}^{R}
    \frac{rdr}{r^\alpha} \: d\phi + \int_{-\frac{\pi}{2}}^{\frac{\pi}{2}}
    \int_{\frac{2R_0+\epsilon_p}{\cos(\phi)}}^{R}
    \frac{rdr}{r^\alpha} \: d\phi \right) \nonumber \\
  &=& \frac{P\lambda}{\alpha-2} \int_{-\frac{\pi}{2}}^{\frac{\pi}{2}}\left(
    \frac{\cos^{\alpha-2}(\phi)}{\epsilon_p^{\alpha-2}}
    + \frac{\cos^{\alpha-2}(\phi)}{(2R_0+\epsilon_p)^{\alpha-2}}    
    - \frac{1}{R^{\alpha-2}} \right) d\phi
\ea
Denote
\be
  A(\alpha) = \int_{-\frac{\pi}{2}}^{\frac{\pi}{2}} \cos^{\alpha-2}(\phi) \: d\phi.
\ee
For an integer $\alpha$, we can compute $A(\alpha)$ in closed
form. We demonstrate a table for some values of $A(\alpha)$ in the
Appendix, which we use in simulations. For other $\alpha$, numerical
evaluation of $A(\alpha)$ is possible. We now can write the second
lower bound on the average interference as
\ba
  E[I_0]_\text{LB2} &=& \frac{P\lambda}{\alpha-2} \left(
    \frac{A(\alpha)}{\epsilon_p^{\alpha-2}}
    + \frac{A(\alpha)}{(2R_0+\epsilon_p)^{\alpha-2}}    
    - \frac{\pi}{R^{\alpha-2}} \right).
\ea
When $R\to\infty$, this lower bound approaches
\ba
  E[I_0]_{LB2}^{\infty} &=& \frac{P\lambda A(\alpha)}{\alpha-2} \left(
    \frac{1}{\epsilon_p^{\alpha-2}} +
    \frac{1}{(2R_0+\epsilon_p)^{\alpha-2}}\right). \label{I0_lower2_Rinfty}
\ea
Since this bound takes into account the interfering transmitters close
to the primary receiver, for a small $\epsilon_p$ or large $R_0$, this
lower bound is tighter than the previous one in (\ref{I0_lower1_Rinfty}).

\subsubsection{An upper bound on $E[I_0]$}

\begin{figure}
\centering{\epsfig{figure=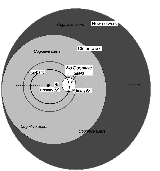, width=7cm}}
\caption{An upper bound on the expected interference at the primary Rx
  is obtained by forming a cognitive-free circle of radius
  $\epsilon_p$ around the primary receiver and enlarging the network
  radius, centered at the primary receiver, to $R+R_0$. All cognitive
  transmitters now lie within these new boundaries.}
\label{fig:upper_bound}
\end{figure}

For the upper bound, similar to the fist lower bound, we re-center the
network at the primary receiver. We now reduce the exclusive region
radius, centered at Rx$_p^1$, to $\epsilon_p$ and extend the outer
network radius, also centered at Rx$_p^1$, to $R_0+R$, as in Figure 
\ref{fig:upper_bound}. The set of cognitive transmitters contained
within these two new circles is a superset of the original, creating an
upper bound on the interference as
\ba
  E[I_0]_\text{UB} &=& \int_{\epsilon_p}^{R_0+R} \frac{2\pi P\lambda
    r}{r^\alpha} dr \nonumber 
  = \frac{2\pi P\lambda}{\alpha-2}
  \left(\frac{1}{\epsilon_p^{\alpha-2}} - \frac{1}{(R+R_0)^{\alpha-2}} 
  \right). \label{I0_upper2}
\ea
As $R\rightarrow\infty$, this upper bound becomes
\ba
  E[I_0]_{U}^{\infty} &=& \frac{2\pi P\lambda}{\alpha-2}
  \frac{1}{\epsilon_p^{\alpha-2}} \label{I0_upper_Rinfty} 
\ea

\subsection{Comparisons of the bounds on the expected interference power}

\begin{figure}
\centering{\epsfig{figure=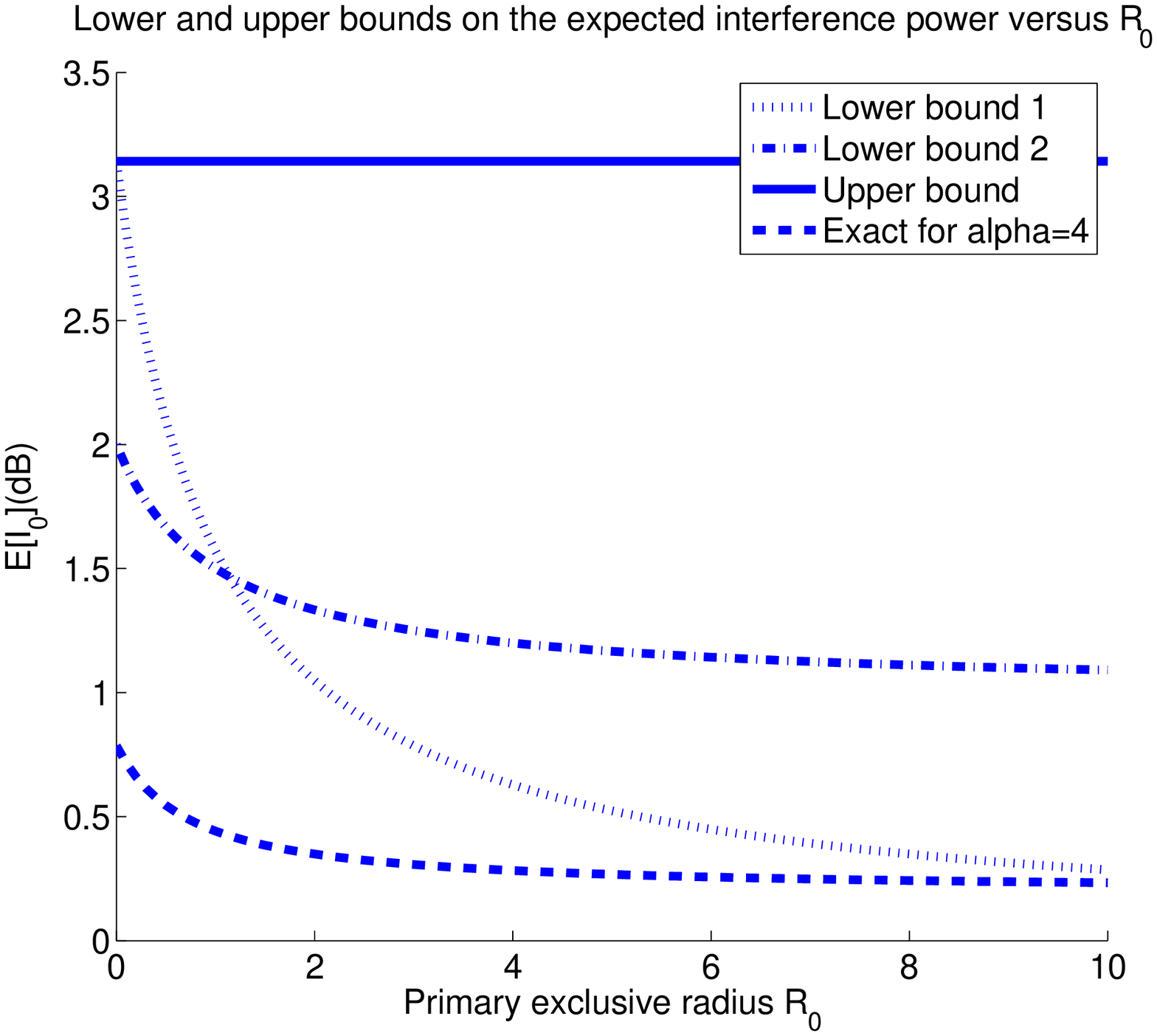, width=8cm}}
\caption{Upper \eqref{I0_upper_Rinfty}, lower bound 1
  \eqref{I0_lower1_Rinfty}, lower bound 2  \eqref{I0_lower2_Rinfty}
  for $\alpha=3$, $\lambda=1$, $P=1$, $\epsilon_p=2$. In this case the
  exact expression for $\alpha=4$ is a lower bound on the expected
  interference.} 
\label{fig:a3e}
\end{figure}

\begin{figure}
\centering{\epsfig{figure=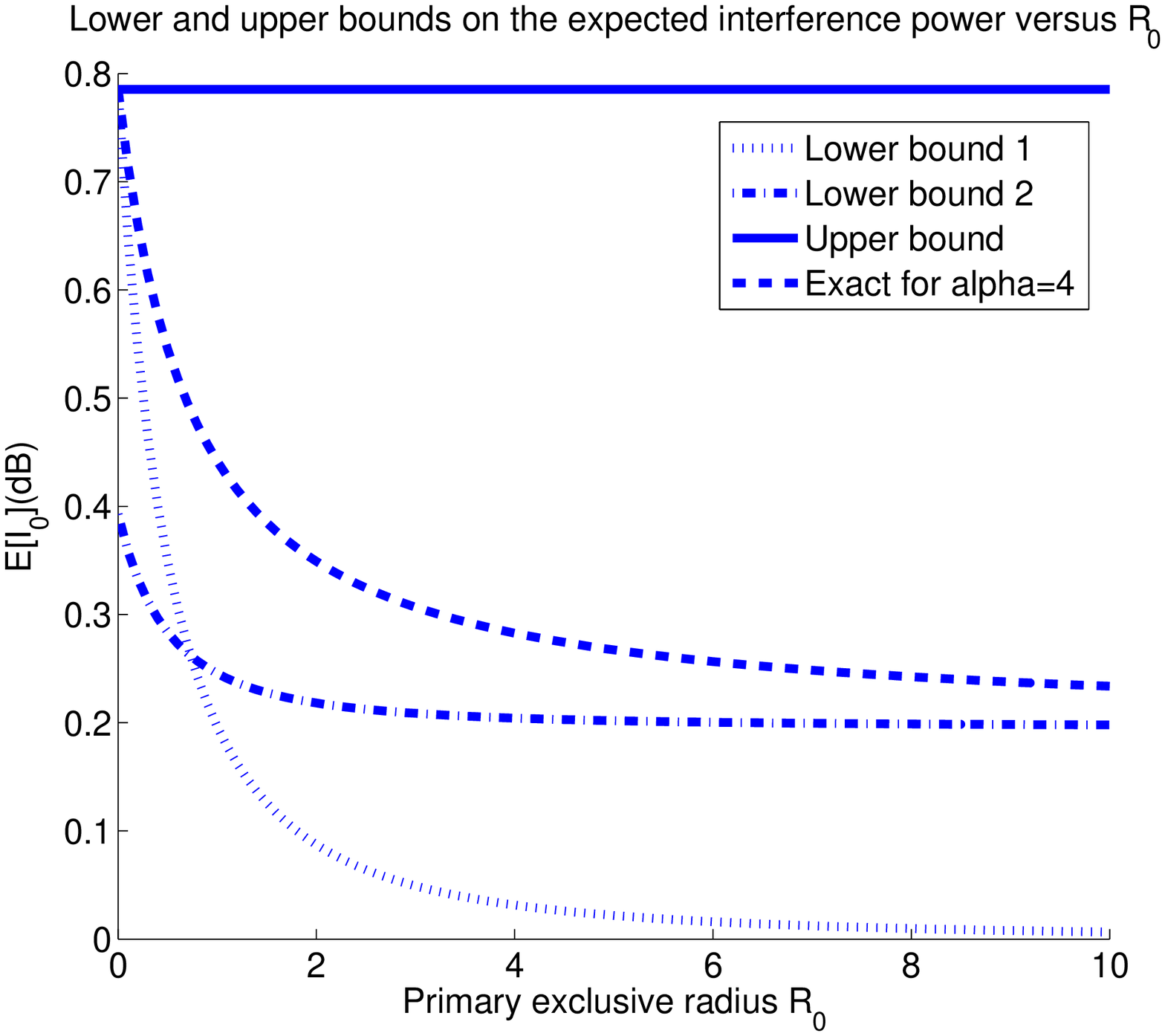, width=8cm}}
\caption{Upper \eqref{I0_upper_Rinfty}, lower bound 1
  \eqref{I0_lower1_Rinfty}, lower bound 2  \eqref{I0_lower2_Rinfty}
  for $\alpha=4$, $\lambda=1$, $P=1$, $\epsilon_p=2$. In this case we
  have the exact expression for $\alpha=4$, which we compare to the
  other bounds to give an indication of their tightness.} 
\label{fig:a4e}
\end{figure}

\begin{figure}
\centering{\epsfig{figure=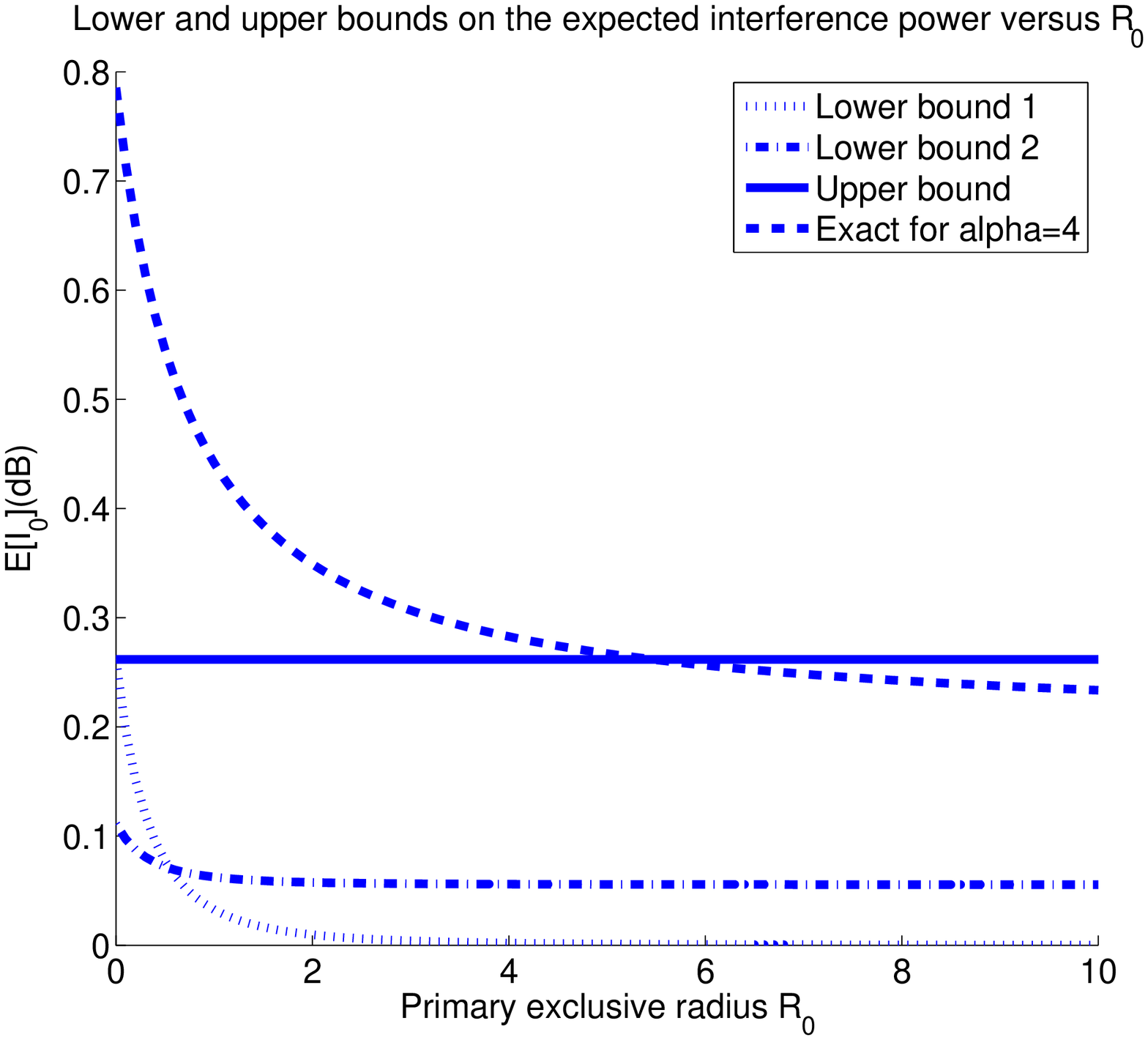, width=8cm}}
\caption{Upper \eqref{I0_upper_Rinfty}, lower bound 1
  \eqref{I0_lower1_Rinfty}, lower bound 2  \eqref{I0_lower2_Rinfty}
  for $\alpha=5$, $\lambda=1$, $P=1$, $\epsilon_p=2$. In this case the
  exact expression for $\alpha=4$ is an upper bound on the expected
  interference.} 
\label{fig:a5e}
\end{figure}

We now compare the upper bound in \eqref{I0_upper_Rinfty} and the
lower bounds in \eqref{I0_lower1_Rinfty} and \eqref{I0_lower2_Rinfty}
for various values of $R_0$ and $\alpha$, while fixing $\lambda=1,
P=1$, and $\epsilon_p=2$ and assuming an infinite network
($R\to\infty$). For $\alpha=3$, Figure \ref{fig:a3e} shows that for
small $R_0$, lower bound 1 is better than lower bound 2, as
expected. The exact expression for the expected interference for 
$\alpha=4$ in (\ref{I0alpha4infty}) provides a lower bound on the
interference for $\alpha=3$. 
For $\alpha=4$, Figure \ref{fig:a4e} shows the upper and lower bounds 
compared to the exact expression of \eqref{I0alpha4infty}. We see that
lower bound 2 is asymptotically tight as $R_0\rightarrow \infty$. For
$\alpha=5$, Figure \ref{fig:a5e} shows the upper and lower bounds. In
this case, the exact expected interference expression 
for $\alpha=4$ yields an upper bound on the expected interference,
which is tighter than the upper bound in \eqref{I0_upper_Rinfty} for
large $R_0$.

\subsection{The primary exclusive radius}

The above bounds on the expected interference can be used to bound the 
radius $R_0$ of the primary exclusive region. In particular, for a
given outage capacity $C_0$, the primary outage constraint
\eqref{outage_constr} can be written as
\ban
  P_e &=& \text{Pr}
  \left[\log_2\left(1+\frac{P_0/R_0^\alpha}{I_0+\sigma^2} 
  \right)\leq C_0 \right] \\
  &=& \text{Pr} \left[I_0 \geq
  \frac{P_0/R_0^\alpha}{(2^{C_0}-1)}-\sigma^2 \right].
\ean
Assuming that the primary network operates in the region that there is
no outage due to noise, then 
\ba
  && \frac{P_0/R_0^\alpha}{(2^{C_0}-1)}-\sigma^2 \geq 0 \nonumber \\
  &\leftrightarrow& R_0 \leq \left( \frac{P_0}{\sigma^2(2^{C_0}-1)}
  \right) ^{1/\alpha} \stackrel{\triangle}{=} R_0^u . \label{R_0_upper_bnd_1}
\ea
If $R_0$ is larger than $R_0^u$, the receivers at the edge of the PER
will be in outage because of noise alone. Thus $R_0^u$ is the maximum
radius to ensure that the outage constraint holds even without any
cognitive users.

Assuming that $R_0$ satisfies (\ref{R_0_upper_bnd_1}), we can apply
Markov's inequality to bound the outage probability as
\ban
  P_e
  &\leq& \frac{E[I_0]}{\frac{P_0/R_0^\alpha}{(2^{C_0}-1)}-\sigma^2 }.
\ean
Assuming an infinite network ($R\to\infty$), using the upper bound on
$E[I_0]$ in (\ref{I0_upper_Rinfty}), we can further bound $P_e$ as
\ban
  P_e &\leq& \frac{2\pi P\lambda}{\alpha-2}
  \frac{1}{\epsilon_p^{\alpha-2}} \left(
    \frac{P_0/R_0^\alpha}{(2^{C_0}-1)}-\sigma^2 \right)^{-1}.
\ean
Bounding this probability by the outage constraint $\beta$, we get
\be
  R_0^\alpha \leq \frac{P_0}{(2^{C_0}-1)} \left( \frac{2\pi
      P\lambda}{\beta(\alpha-2)} \frac{1}{\epsilon_p^{\alpha-2}} +
    \sigma^2 \right)^{-1} .
  \label{R_0_upper_bnd_2}
\ee
This bound is always smaller than the bound in
(\ref{R_0_upper_bnd_1}). Thus, as expected, the maximum distance that
we can guarantee an outage probability for a primary receiver will be
reduced in the presence of cognitive users.

When $\alpha$ is an even integer, we can use the exact value of
$E[I_0]$ in the Markov inequality to obtain a tighter bound on
$R_0$. Using the example for $\alpha = 4$ in (\ref{I0alpha4infty}), 
we obtain an implicit equation for all exclusive region radii $R_0$
such that \eqref{outage_constr} holds as 
\begin{equation}
\frac{(R_0+\epsilon_p)^2}{\epsilon_p^2(2R_0+\epsilon_p)^2}\leq
\frac{\beta}{\lambda \pi P }\left(
\frac{P_0/R_0^4}{2^{C_0}-1}-\sigma^2\right). \label{eq:implicit} 
\end{equation}

Equations (\ref{R_0_upper_bnd_2}) and (\ref{eq:implicit}) provide a
relation among the system parameters: $P_0$ (the primary transmit
power), $P$ (the cognitive users' power), $C_0$ (the outage capacity),
$\beta$ (the outage probability), $\lambda$ (the cognitive user
density), $\sigma^2$ (the noise power), and $R_0$ (the exclusive
region radius). These equations may be of particular interest when
designing the primary system to guarantee the primary outage
constraint $\Pr[\text{primary user's rate} \leq C_0]\leq \beta$. By
fixing several of the parameters, we can obtain relations among the
others. Specifically, we relate the primary outage target rate $C_0$
to the capacity without interference $C = \log_2(1+P_0/\sigma^2)$ as
$C_0 = \eta C$, where $0 \leq \eta \leq 1$ represents the fraction of
the interference-free capacity that we wish to guarantee with
probability $\beta$ in a sea of cognitive users.
 
\begin{figure}
\centerline{\epsfig{figure=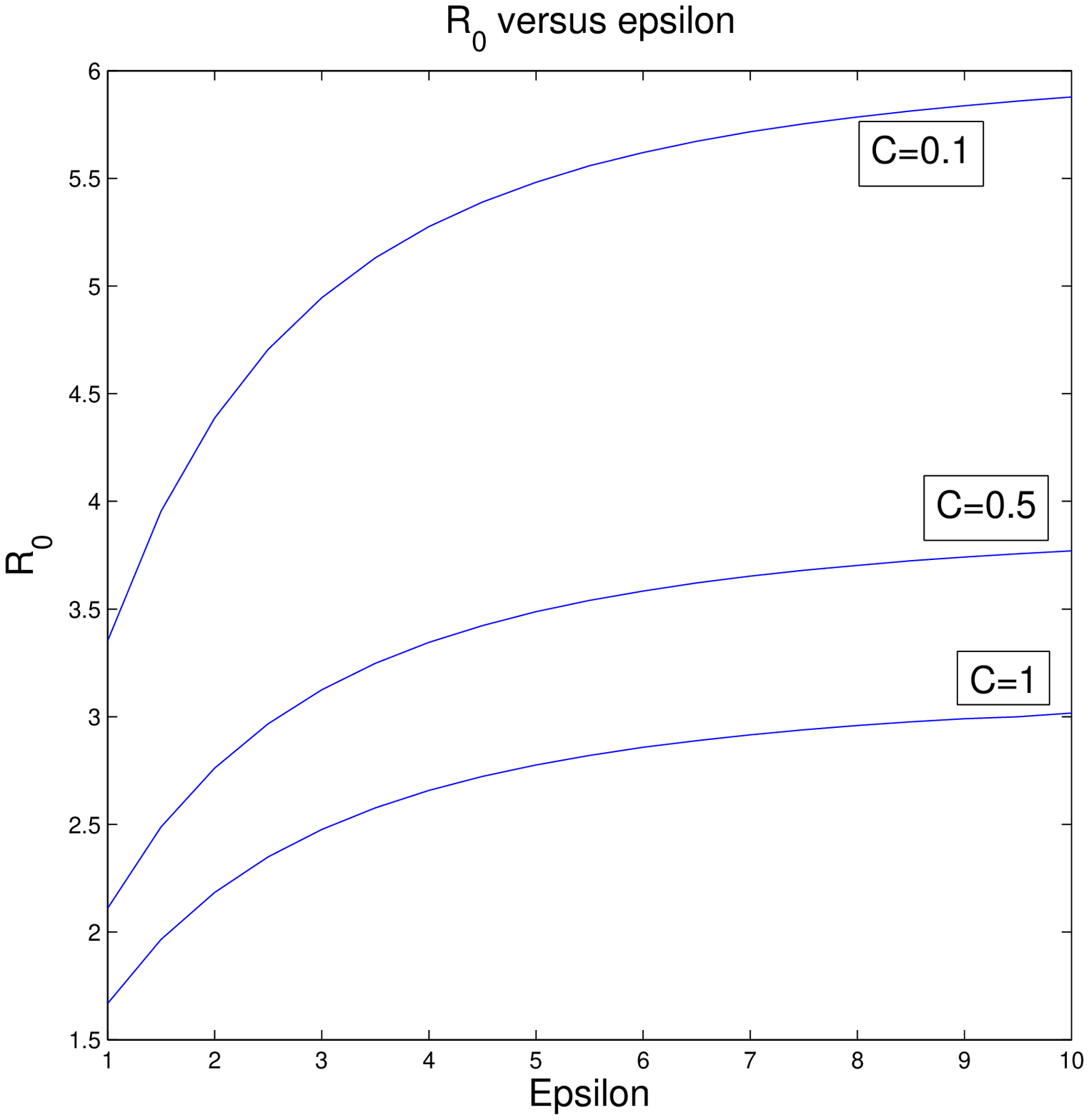, width=8cm}}
\caption{The relation between the exclusive region radius $R_0$ and
  the guard band $\epsilon_p$ according to \eqref{R_0_upper_bnd_2} for 
  $\lambda=1, P=1, P_0=100, \sigma^2=1, \beta=0.1$ and $\alpha=3$.}
\label{fig:R0_e_C0}
\end{figure}

As an example, we plot in Figure \ref{fig:R0_e_C0} the relation
between the exclusive region radius $R_0$ and the guard-band width
$\epsilon_p$ for various values of the outage capacity $C_0$, while
fixing all other parameters according to (\ref{eq:implicit}) for
$\alpha=4$. The plots show that $R_0$ increases with $\epsilon_p$, and
the two are of approximately the same order. This is intuitive since
at the primary receiver there is a trade-off between the interference
seen from the secondary users, which is of a minimum distance
$\epsilon_p$ away, and the desired signal strength from the primary
BS, which is of the distance $R_0$ away. The larger the $\epsilon_p$,
the less interference, and thus the further away the primary receiver
may lie from the base station. We also notice that as $C_0$ increases,
$R_0$ decreases for the same $\epsilon_p$. This is again intuitive: as
we require a higher capacity, the relative interference (to the
desired signal) must be reduced, which is achieved by reducing $R_0$
for a fixed $\epsilon_p$. Finally, as $\epsilon_p\rightarrow\infty$,
$R_0$ approaches the limit of the interference-free bound in
(\ref{R_0_upper_bnd_1}) for $\alpha = 4$.

\begin{figure}
\centerline{\epsfig{figure=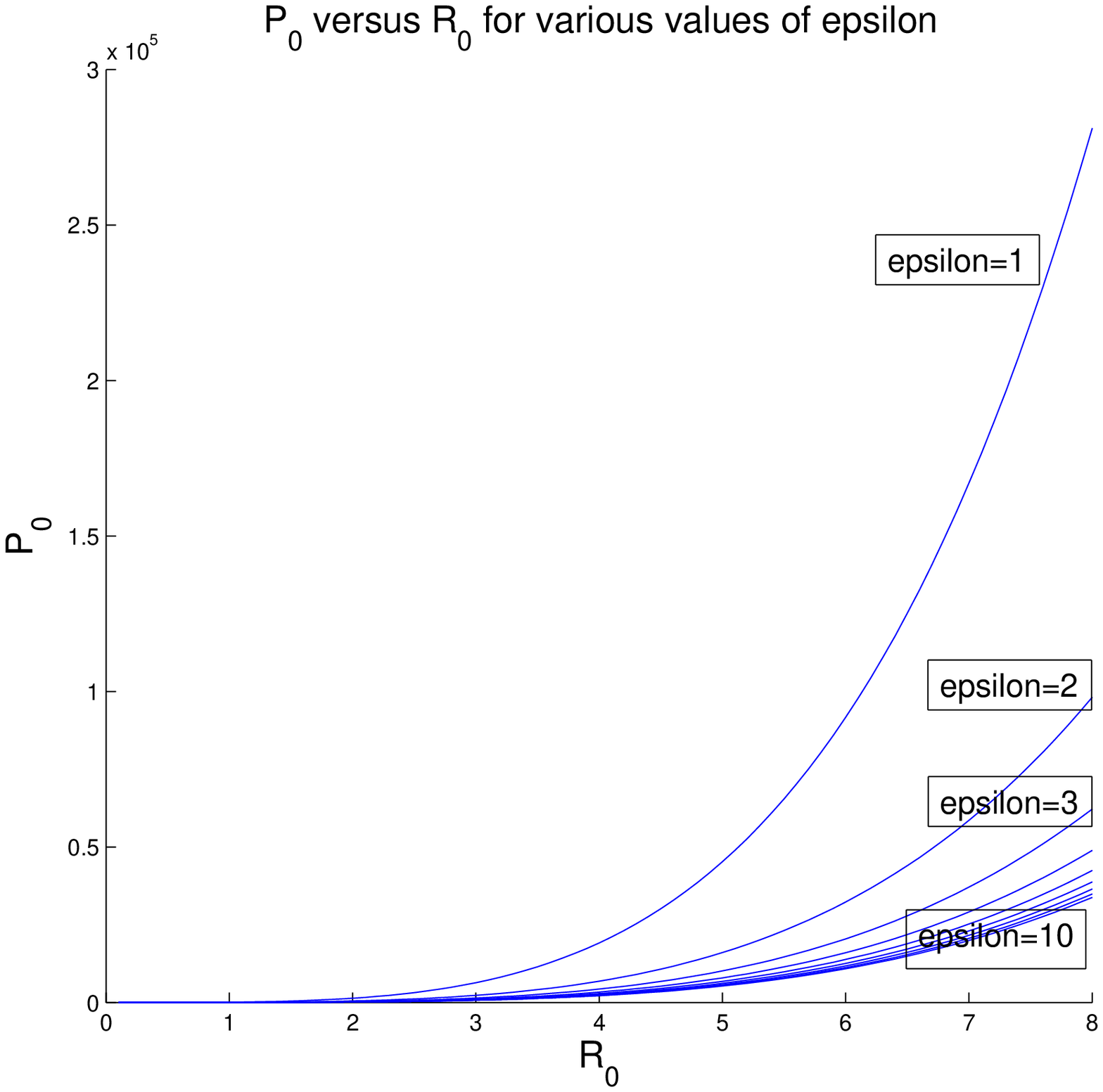, width=8cm}}
\caption{The relation between the BS power $P_0$ and the  exclusive
  region radius $R_0$ according to \eqref{R_0_upper_bnd_2} for
  $\lambda=1, P=1, \sigma^2=1, \beta=0.1, C_0=3$ and $\alpha=3$.}
\label{fig:P0_R0_e}
\end{figure}

Alternatively, we can fix the guard band $\epsilon_p$ and the
secondary user power $P$ and seek the relation between the primary 
power $P_0$ and the exclusive radius $R_0$ that can support the outage
capacity $C_0$. In Figure \ref{fig:P0_R0_e}, we plot this relation
according to (\ref{eq:implicit}) for $\alpha = 4$.
The fourth-order increase in power here is inline with the path loss
$\alpha=4$. Interestingly, a small increase in the gap band
$\epsilon_p$ can lead to a large reduction in the required primary
transmit power $P_0$ to reach a receiver at a given radius $R_0$ 
while satisfying the given outage constraint.

\section{Single primary user network with 
cognitive distance-dependent power scaling}
\label{sec:ps}
 
In this section, we consider the case in which cognitive transmitters
can scale their power according to the distance to a single primary user
located at the center of the network. Such a model is relevant when
primary transmitters are spread out, or whenever the interference from
other primary users is negligible. Suppose that the network consists
of a single primary user with the transmitter at the center and the
receiver within a radius $R_0$, where we assume $R_0 > 1$ in this
section. Then, intuitively the cognitive transmitters further away
from the center may transmit at a higher power without significantly
increasing the interference to the primary receiver. We confirm that
this cognitive power scaling does not affect the cognitive user
scaling law. Furthermore, it allows $D_{max}$, the maximum distance
between a cognitive transmitter and receiver, to grow with the network
size. We also explore the impact of distance-dependent cognitive power
scaling on the expected interference at the primary receiver and on
the PER radius $R_0$.

Assume the transmit power of a cognitive user at radius $r$ is
\be
  P = P_c r^\gamma
  \label{cog_power_scale}
\ee
where $P_c$ is a constant, and $\gamma$ is the cognitive power
exponent. As shown later, for the interference from the cognitive
users to stay bounded, we require that $\gamma < \alpha -2$.

\subsection{The effect of cognitive power scaling on $D_{max}$}

Let the maximum distance between a cognitive Tx and Rx be
$D_{\max}$. Then  the cognitive channel gain is given by $|h_{ii}|^2
\geq 1/D_{\max}^\alpha$. The received power at the cognitive receiver
will be
\be
  P_r = \frac{P}{|h_{ii}|^2} \leq \frac{P_c r^\gamma}{D_{\max}^\alpha}.
  \label{rx_power_bnd}
\ee
Consider the ratio $\frac{r^\gamma}{D_{\max}^\alpha}$. If we constrain
this ratio to be lower-bounded by a constant, then $D_{\max}$ can grow
with distance such that 
\be
  D_{\max} \leq K_d r^{\gamma/\alpha} .
  \label{Dmax_bound}
\ee
for some constant value $K_d$. Therefore, when we let the transmit
power of cognitive users scale with distance, then the maximum
distance between a cognitive Tx and Rx can also scale with
distance. Noting that $\gamma < \alpha - 2$, and since $r \geq R_0 +
\epsilon_p > 1$, we have 
\be
  D_{\max} \leq K_d r^{\gamma/\alpha} < K_d r^{1-2/\alpha}.
\ee
Thus depending on the path loss $\alpha$, the cognitive Tx-Rx distance
can grow with an exponent of upto $1-2/\alpha$. For a large $\alpha$,
this growth is almost at the same rate as the network.

\subsection{Cognitive throughput scaling}

We  first examine the throughput scaling of the cognitive users as
$n\to\infty$. Consider a cognitive receiver, the interference from the
single primary user will be bounded due to the finite transmit
power and the exclusive radius $R_0$. The interference from the other
cognitive users, however, is different.

\subsubsection{Interference from other cognitive users}
The scaled cognitive transmit power affects the interference to a
cognitive receiver from other cognitive users. Similar to the
development in (\ref{avg_cog_interf_1}) and (\ref{avg_cog_interf_n}),
this average interference in the worst-case is now upper bounded by
\be
  I_\text{avg, n} = 2\pi\lambda \int_{\epsilon_c}^{R} r r^\gamma
  r^{-\alpha} dr = \frac{2\pi\lambda}{\alpha-2-\gamma} \left(
  \frac{1}{\epsilon_c^{\alpha-2-\gamma}} - \frac{1}{R^{\alpha-2-\gamma}}
  \right).
\ee
If $\gamma < \alpha - 2$, then this average interference is
bounded as $R\to\infty$. Thus we can let the cognitive users transmit
with higher power the further they are from the primary user, as long
as their power scaling satisfies $\gamma < \alpha - 2$. This power
scaling therefore is dependent on the propagation environment.

With $\gamma < \alpha - 2$, as $n\to\infty$, the average interference
from other cognitive users approaches
\be
  I_{\infty}^{(\gamma)} = \frac{2\pi\lambda}{\alpha-2-\gamma}
  \frac{1}{\epsilon_c^{\alpha-2-\gamma}} .
  \label{I_avg_infty_gamma}
\ee
This can be seen as the interference with an effective path loss
decreased to $\alpha - \gamma$. Therefore, the scaling in transmit
power of cognitive users takes the advantage of the margin in a high
path loss when signal power decays faster than 2.

\subsubsection{Lower bound on the cognitive network throughput}
The rate of a cognitive user (\ref{cog_rate}) can now be written as
\be
  C_i = \log\left( 1 + \frac{P_c r^\gamma |h_{ii}|^2}
  {I_{pi} + \sigma^2 + I_{ci}} \right).
\ee
Recall that $I_{pi} \leq I_P$ (\ref{I_P}). Together with the bounds on
$D_{\max}$ in (\ref{Dmax_bound}) and on the receive power in
(\ref{rx_power_bnd}), we now have
\be
  C_i \geq \log\left( 1 + \frac{P_c}
  {(I_P + \sigma^2 + I_{ci}) K_d^\alpha } \right)
\ee
Again applying the Jensen inequality, the average rate of each
cognitive user satisfies
\ba
  E[C_i] &\geq& \log\left( 1 + \frac{P_c}
  {(I_P + \sigma^2 + E[I_i]) K_d^\alpha } \right) \nonumber 
\ea
As $n\to\infty$, the lower bound approaches a constant as
\be
  E[C_i] \geq \log\left( 1 + \frac{P_{r,\min}} {(I_P + \sigma^2 +
   I_{\infty}^{(\gamma)}) K_d^\alpha } \right) 
  \stackrel{\triangle}{=} \bar{C}_1^{(\gamma)}
  \label{rate_lower_bnd_gamma}
\ee
where $I_{\infty}^{(\gamma)}$ is given in
(\ref{I_avg_infty_gamma}). Again the average per-user throughput of
this cognitive network remains at least a constant as
$n\to\infty$. Applying a concentration analysis similar to Section
\ref{sec:concent}, we conclude that with cognitive transmit power scaling
(\ref{cog_power_scale}), the sum throughput of the cognitive users in
any network also scales linearly with the number of users.

\subsection{Effect of cognitive power scaling on the interference at
  the primary receiver} 

Similar to Section \ref{sec:pe}, we examine the effect on the 
expected interference $E[I_0]$ at the primary receiver of having
cognitive transmitters scale their power according to the distance
from the primary user as in (\ref{cog_power_scale}). This expected
interference may now be expressed as
\begin{align}
  E[I_0] &= n \int_{R_0+\epsilon_p}^{R} \int_0^{2\pi} \frac{P_c r^{\gamma}}{
    d(r,\theta)^{\alpha}} f_r(r) f_\theta(\theta) \; dr \; d\theta \nonumber \\
  &= \int_{R_0+\epsilon_p}^R  \int_0^{2\pi}
  \frac{\lambda r^{\gamma+1} P_c \; dr\; d\theta}{(r^2 + R_0^2 - 2R_0 r
    \cos\theta)^{\alpha/2}}. \label{eq:E0gamma}
\end{align}

\medskip

We notice the additional factor $r^{\gamma}$ in the
numerator. Similarly, the two lower bounds and single upper bound
derived in Section \ref{sec:pe} may also be changed to reflect the
power scaling. The bounds, in terms of $\gamma$, may be expressed as
\eqref{eq:gammalb1}, \eqref{eq:gammalb2} and \eqref{eq:gammaup}, where
we require that $\gamma<\alpha-2$, and $R\rightarrow \infty$ for
simplicity.

\begin{align}
E[I_0]_{LB 1}^{\infty}(\gamma) &= \frac{2\pi P_c\lambda}{\alpha-2-\gamma}
  \frac{1}{(2R_0+\epsilon_p)^{\alpha-2-\gamma}} \label{eq:gammalb1}\\
    E[I_0]_{LB2}^{\infty}(\gamma) &= \frac{P_c\lambda A(\alpha-\gamma)}{\alpha-2-\gamma} \left(
    \frac{1}{\epsilon_p^{\alpha-2-\gamma}} +
    \frac{1}{(2R_0+\epsilon_p)^{\alpha-2-\gamma}}\right) \label{eq:gammalb2} \\
 E[I_0]_{U}^{\infty}(\gamma) & = \frac{2\pi P_c\lambda}{\alpha-2-\gamma}
  \frac{1}{\epsilon_p^{\alpha-2-\gamma}} \label{eq:gammaup}
\end{align}

These bounds may be interpreted as follows. For a given path loss
$\alpha$ and acceptable power scaling of $\gamma$ (such that the
cognitive users may achieve the same linear scaling law as when power
scaling was not employed), these bounds correspond to those of a
channel with no power scaling and path loss $\alpha^* =
\alpha-\gamma$. Again, a network with power scaling may be thought
of as an equivalent network without power scaling but with a slower 
decay of the power with distance (a smaller path loss parameter). As
an example, for $\alpha = 5$, the plots of bounds in Figure
\ref{fig:a5e} apply when $\gamma = 0$, in Figure \ref{fig:a4e} when
$\gamma = 1$ and in Figure \ref{fig:a3e} when $\gamma=2$.


\section{Conclusion}
\label{sec:cn}

As secondary spectrum usage is rapidly approaching, it is important to 
study the potential of cognitive radios and cognitive transmission
from a \emph{network} perspective. In this paper, we have determined
the sum-rate scaling of a network of one-hop cognitive
transmitter-receiver pairs which simultaneously communicate, while   
probabilistically guaranteeing the primary user link a minimum rate.
With simultaneous one-hop cognitive transmissions, we show that the
sum-rate of cognitive users scales linearly in the number of cognitive
links $n$ as $n\rightarrow \infty$. This result holds in presence of
multiple primary users, when the cognitive transmitters use constant
power. The same result also holds in the presence of a single primary
user, when the cognitive transmitters scale their power according
to the distance from the primary user. Then using the outage
constraint on the primary user, we derive bounds on the radius of a 
primary exclusive region (PER) around each primary transmitter. These
bounds help in the design of a primary network with PERs such that, 
outside these regions, uniformly distributed cognitive transmitters
may freely transmit while not harming the primary user.

\fontsize{11}{11}
\selectfont
\bibliographystyle{IEEEtran}
\bibliography{jstsp_cognet}


\begin{appendix}

\section*{Bounds on the interference from the primary users}

\medskip
To show that $I_A$ in (\ref{I_A}) is bounded, we use the following inequalities:
\ban
  \frac{1}{(x^2+y^2)^{\alpha/2}} \leq
  \frac{2^{\alpha/2}}{(x+y)^\alpha} \;, \quad x+y > 0
\ean
and
\ban
  \sum_{k=0}^\infty \frac{1}{(ak + b)^{\alpha}} & \leq &
  \frac{1}{b^\alpha} + \int_0^\infty \frac{1}{(ax + b)^{\alpha}} dx \;,
  \quad b > 0 \\
  &=& \frac{1}{b^\alpha} + \frac{1}{(\alpha - 1)a}
  \frac{1}{b^{\alpha-1}} \;, \quad b > 0 .
\ean
Applying these inequalities to the following generic sum with $a>0$,
$c>0$ and $b+d > 0$ as:
\ba
  && \sum_{m=0}^\infty \sum_{k=0}^\infty
  \frac{1}{\left[(ak+b)^2 + (cm+d)^2\right]^{\alpha/2}} \label{gen_sum}\\
  &\leq& \sum_{m=0}^\infty \sum_{k=0}^\infty
  \frac{2^{\alpha/2}}{\left(ak+b + cm+d\right)^{\alpha}} \nonumber\\
  &\leq& 2^{\alpha/2} \sum_{m=0}^\infty
  \left(\frac{1}{\left(c m + {b+d}\right)^\alpha} +
  \frac{1}{(\alpha-1)a}\frac{1}{\left(c m + {b+d}\right)^{\alpha-1}}
  \right) \nonumber\\
  &\leq& 2^{\alpha/2} \left[ \frac{1}{\left({b+d}\right)^\alpha} +
  \frac{1}{(\alpha-1)c}\frac{1}{\left({b+d}\right)^{\alpha-1}} + 
  \frac{1}{(\alpha-1)a}\left( \frac{1}{\left({b+d}\right)^{\alpha-1}}
  + \frac{1}{(\alpha-2)c}\frac{1}{\left({b+d}\right)^{\alpha-2}}
  \right) \right] .\nonumber
\ea
Thus for $\alpha>2$, this summation is bounded.

Now consider $I_A$ in (\ref{I_A}). Denote $I_{A1}$ as the first
double-summation, then it can be rewritten as
\ban
  I_{A1} 
  &=& \sum_{m=0}^\infty \sum_{k=0}^\infty
  \frac{1}{\left[\left(2\sqrt{3}k-\epsilon_c\cos\theta\right)^2 + 
      \left(2m-\epsilon_c\sin\theta\right)^2\right]^{\alpha/2}} \\
  &+& \sum_{m=0}^\infty \sum_{k=0}^\infty
  \frac{1}{\left[\left(2\sqrt{3}k+\epsilon_c\cos\theta\right)^2 + 
      \left(2m+\epsilon_c\sin\theta\right)^2\right]^{\alpha/2}} -
  \frac{1}{\epsilon_c^\alpha}
\ean
Since $\epsilon_c < 1$ (normalized to $R_0$), then
$|\epsilon_c\cos\theta| \leq 1$ and $|\epsilon_c\sin\theta| \leq 1$
for all $\theta\in[0,2\pi]$. For each of the double-summations in
$I_{A1}$, after separating out the first three finite terms
corresponding to $(k=0,m=0)$, $(k=0,m=1)$, and $(k=1,m=0)$, then the
rest can be rewritten in the form (\ref{gen_sum}) with $b+d>0$, which
is bounded. Similarly for the second summation in $I_A$
(\ref{I_A}). Therefore for any $\theta$, $I_A$ (\ref{I_A}) is 
bounded.


\section*{Calculation of the exact $E[I_0]$ when $\alpha=4$}

For $a>|b|$, from pg. 383 \cite{integrals}, we obtain
\[ \int_0^{2\pi}\frac{dx}{(a+b\cos(x))^{2}} = \frac{2\pi a}{(a^2-b^2)^{3/2}}\]

In the integral of interest \eqref{eq:int1} we have $a=R_0^2+r^2$ and
$b=-2R_0r$, and so $R_0^2+r^2> 2R_0 r$ as needed. Thus, the expected
interference from all cognitive users is given by \eqref{eq:av5}.
\begin{align}
 E[I_0] &= \lambda \pi P  \int_{R_0+\epsilon_p}^{R} \int_{0}^{2\pi}\frac{2r\; dr
   \; d\theta}{2\pi (R_0^2+r^2-2R_0 r \cos\theta)^{2}} \notag 
= \lambda \pi P \int_{R_0+\epsilon_p}^{R} \frac{2r(r^2+R_0^2)}{(r^2-R_0^2)^3} \; dr \notag \\
&= \lambda \pi P
 \left.
 \left[-\frac{r^2+R_0^2}{2(r^2-R_0^2)^2}-\frac{1}{2(r^2-R_0^2)}\right]
 \right|_{R_0+\epsilon_p}^{R} \notag \\  
&= \lambda \pi P
 \left[-\frac{R^2}{(R^2-R_0^2)^2}+\frac{(R_0+\epsilon_p)^2}{\epsilon_p^2(2R_0+\epsilon_p)^2}
 \right]  \label{eq:av5} 
\end{align}

Thus, if we let the number of users $n\rightarrow\infty$, or
equivalently, as $R\rightarrow \infty$, the total interference
experienced by the primary receiver when on the edge of the primary
exclusive region approaches the constant
\[ E[I_0]_{\infty} = \frac{\lambda \pi P (R_0+\epsilon_p)^2}{\epsilon_p^2(2R_0+\epsilon_p)^2}.\]


\section*{Evaluation of $A(\alpha)$}

The lower bounds  on the expected value and variance of the
interference depend on the function
\[ A(\alpha) = \int_{-\frac{\pi}{2}}^{\frac{\pi}{2}} \cos^{\alpha-2}(\phi) \; d\phi \]
This function may be easily  calculated (see for example pg. 161 of
\cite{integrals}) for integral values of $\alpha$. For completeness,
and reference for our simulations, here is a table of $A(\alpha)$. 

\bigskip

\begin{center}
\begin{tabular}{|c|ccccccccc|}
\hline
$\alpha$ & 2 & 3 & 4 & 5 & 6 & 7 & 8 & 9 & 10 \\
\hline
$A(\alpha)$ & $\pi$
& 2  & $\pi/2$ & 4/3 & $3\pi/8$ & 16/15 & $5\pi/16$ & 32/35 & $35\pi/128$\\
\hline
\end{tabular}
\end{center}

\end{appendix}

\end{document}